\documentclass[12pt]{article}%

\usepackage[utf8]{inputenc}
\usepackage[T1]{fontenc}
\usepackage{textcomp}
\usepackage{gensymb}
\usepackage{amsmath}
\usepackage[margin=1.12in]{geometry}
\usepackage{graphicx}
\usepackage{amssymb}
\usepackage{authblk}
\usepackage{blindtext}
\usepackage{apacite}
\allowdisplaybreaks
\usepackage{graphics}
\usepackage{float}
\usepackage{setspace}
\usepackage{amsthm}
\theoremstyle{plain}

\usepackage{algorithm}
\usepackage{algpseudocode}

\newcommand{\bpsi}{ \mbox{\boldmath $ \psi$} }
\newcommand{\bPsi}{ \mbox{\boldmath $ \Psi$} }

\newcommand{\bbeta}{ \mbox{\boldmath $ \beta $} }

\newcommand{\bphi}{ \mbox{\boldmath $\phi$}}

\newcommand{\bzero}{\textbf{0}}
\newcommand{\bone}{\textbf{1}}

\newcommand{\bz}{\textbf{z}}
\newcommand{\ba}{\textbf{a}}

\newcommand{\bB}{\textbf{B}}

\newcommand{\bD}{\textbf{D}}

\newcommand{\bE}{\textbf{E}}

\newcommand{\bF}{\textbf{F}}

\newcommand{\bh}{\textbf{h}}

\newcommand{\bI}{\textbf{I}}

\newcommand{\bm}{\textbf{m}}

\newcommand{\bs}{\textbf{s}}

\newcommand{\bV}{\textbf{V}}
\newcommand{\bw}{\textbf{w}}

\newcommand{\bx}{\textbf{x}}
\newcommand{\bX}{\textbf{X}}
\newcommand{\by}{\textbf{y}}
\newcommand{\bY}{\textbf{Y}}

\restylefloat{table}
\usepackage{subcaption}
\usepackage{amssymb}
\usepackage{hhline}
\RequirePackage[OT1]{fontenc}
\RequirePackage{amsthm,amsmath}
\RequirePackage{natbib}
\RequirePackage[colorlinks,citecolor=blue,urlcolor=blue]{hyperref}

\begin{document}

\title{Non-separable Nearest-Neighbor Gaussian Process Model for Antarctic Surface Mass Balance and Ice Core Site Selection}
\author[1]{Philip A. White \thanks{Corresponding Author: paw27@duke.edu}}
\author[2]{C. Shane Reese \thanks{reese@stat.byu.edu}}
\author[2]{William F. Christensen \thanks{william@stat.byu.edu}}
\author[3]{ Summer Rupper \thanks{ summer.rupper@geog.utah.edu}}
\affil[1]{Department of Statistical Science, Duke University, Durham, NC, USA}
\affil[2]{Department of Statistics, Brigham Young University, Provo, UT, USA}
\affil[3]{Department of Geography, University of Utah, Salt Lake City, UT, USA}

\maketitle

\begin{abstract}
Surface mass balance (SMB) is an important factor in the estimation of sea level change, and data are collected to estimate models for prediction of SMB over the Antarctic ice sheets. Using a quality-controlled aggregate dataset of SMB field measurements with significantly more observations than previous analyses \citep{favier2013}, a fully Bayesian nearest-neighbor Gaussian process model is posed to estimate Antarctic SMB and propose new field measurement locations. A corresponding Antarctic SMB map is rendered using this model and is compared with previous estimates. A prediction uncertainty map is created to identify regions of high SMB uncertainty. The model estimates net SMB to be 2345 Gton $\text{yr}^{-1}$, with 95\% credible interval (2273,2413) Gton $\text{yr}^{-1}$. Overall, these results suggest lower Antarctic SMB than previously reported. Using the model's uncertainty quantification, we propose 25 new measurement sites for field study utilizing a design to minimize integrated mean squared error.
\end{abstract}

\noindent\textsc{Keywords}: {spatial statistics, Bayesian statistics, covariance functions, climate change, uncertainty quantification, ice sheet
}


\vspace{-3mm}
\section{Introduction}
Antarctica covers an area larger than the combined area of China and India, with about 98\% of that area covered in ice. The volume of Antarctic ice is equivalent to nearly 60m of global sea level rise. Thus, even small changes in the Antarctic ice sheets would have significant impacts on global sea level, as well as associated changes in ocean currents and global climate. To understand Antarctic ice sheet response to climate change, and thus quantify impacts of changes in the Antarctic ice sheets on sea-level rise and climate, the spatial and temporal variability in ice sheet mass balance must first be accurately quantified. Ice sheet mass balance is the difference between the sum of all incoming mass and the sum of all mass lost ($\sum \mbox{incoming} - \sum \mbox{loss}$). If the ice sheet mass balance is negative, then there is a net flux of water to the oceans, and the ice sheet is contributing to sea level rise. The reverse is true if the ice sheet mass balance is positive.

One significant component of ice sheet mass balance is surface mass balance (SMB). As defined here, SMB is the net precipitation, sublimation, melt, and wind redistribution of snow. For most of Antarctica, SMB is positive (net mass gain) and accounts for the incoming mass to the ice sheet. Most mass loss occurs along the margins of the ice sheet via melting under the floating ice shelves and the breaking off of large icebergs from ice sheet margins, a process called calving. Since climatic change affects precipitation, sublimation, melt, and wind over the ice sheet, SMB is directly linked to changes in climate. Thus, more accurate quantification of SMB will greatly improve our understanding of mass balance processes, provide a direct link to climate drivers of ice sheet mass balance and ice sheet dynamics, and provide a reasonable target for climate and ice sheet process models.


Surface mass balance data can be acquired from eclectic methods and sources, including snow stakes, ice cores, satellite altimetry, and radar propagation \citep{magand2007}. For point-wise estimates, SMB is often reported as an average rate of accumulation in units of mm w.e. $\textrm{yr}^{-1}$ (millimeters water equivalents per year), while SMB integrated over large regions is normally given in Gton yr$^{-1}$ (gigatons per year). Because of SMB's importance, researchers traverse Antarctica to install snow stakes, drill ice cores, and dig snow pits for SMB measurements. These SMB field measurements are assumed to be more reliable than remote sensing data; however, not all SMB measurement methods are equally reliable. When SMB measurement method reliability was analyzed, \citet{magand2007} found that long-term ice stakes and ice cores dated with anthropogenic radionuclides are the most reliable direct SMB measurement methods. 

Because data acquisition on Antarctica is expensive, arduous, and restricted to accessible and geophysically appropriate locations, and given Antarctica's immense size, data are unevenly spaced and sparse. For this reason, many models have been developed to estimate SMB in regions lacking data. Interpolative methods based on remote sensing measurements such as passive microwave and laser altimetry have been used by \citet{vaughan1999} to interpolate between 1860 in situ measurements. \citet{vaughan1999} estimated net SMB over the grounded ice sheet as 1811 Gton $\text{yr}^{-1}$ and 2288 Gton $\text{yr}^{-1}$ over all ice sheets (including ice shelves). Although Vaughan et al. discuss uncertainty associated with their model, no error is given for their estimates. Using microwave emission and 540 in situ measurements with 99\% of the data coming between 1950-2000, \citet{arthern2006} predicted net SMB of $1768 \pm 49$ Gton $\text{yr}^{-1}$ over the grounded ice sheet using universal kriging. While their method for determining regional error is given, it is less clear how the uncertainty for SMB is computed. Using a variety of climate models from 1979 to 1999, \citet{bromwich2004} estimate ice sheet SMB to be $2572 \pm 221$ Gton $\text{yr}^{-1}$. Again, it is unclear how the uncertainty in their model is calculated. Calibrating climate model output from the period 1980-2004 to SMB observations and using weighted averages, \citet{vandeberg2006} estimate net SMB on the grounded ice sheet as $2076 \pm 29$ Gton $\text{yr}^{-1}$ and $2521 \pm 29$ Gton $\text{yr}^{-1}$ over the entire ice sheet; then, they used 10,000 model calibrations to obtain uncertainty estimates. In 2012, \citet{lenaerts2012} utilized regional-scale climate models over 1979-2010 to estimate SMB as $2418 \pm 181$ Gton $\text{yr}^{-1}$ over the entire ice sheet. To obtain uncertainty bounds, they used comparisons between their model to SMB observations. While similar, these SMB estimates and associated uncertainties demonstrate how widely Antarctic SMB estimates vary, even when averaged or integrated across the ice sheets. 

Three primary reasons motivate our reanalysis of Antarctic SMB. First, updated data compilations of \citet{favier2013} allow us to use more field measurements ($N=5564$) than have previously been utilized. Second, rigorous uncertainty quantification accounting for spatially correlated errors, measurement reliability, and model parameter uncertainty enable us to more accurately model Antarctic SMB processes. Lastly, identifying new field measurement locations that will reduce future uncertainty in total SMB is of prime importance to glaciologists and climate scientists. Given the scientific community's financial commitment to better characterize the nature of Antarctic climate change, optimal allocation of new data acquisition is an important challenge that is often approached in an ad hoc fashion but is a challenge that we address in a data-based manner. With more rigorous statistical methods and more available data, we can update and refine previous SMB estimates and propose locations for acquiring new data. 

In this paper, we begin by discussing the data characteristics and issues addressed in this analysis in Section \ref{sec:dat}. Section \ref{sec:covar} presents potential covariance models used. In Section \ref{sec:modcomp}, we compare competing models to select a final model used for the remainder of the analysis. Then, in Section \ref{sec:mod}, the statistical model that accounts for spatial correlation and varying levels of reliability in the data is posed. We propose a method for recommending new field measurements that reduces future uncertainty in Antarctic SMB in Section \ref{sec:prop}. Lastly, Sections \ref{sec:res} and \ref{sec:disc} discuss the results of our model and their glaciological implications, as well as compare our results to previous results. 

\vspace{-3mm}
\section{Data}\label{sec:dat}
\begin{figure}[H] 
  \begin{center}
  \scriptsize
    \begin{subfigure}[b]{.32\textwidth}
\begin{tabular}{lrr}
  \hline
 Decade & Count & Percentage \\ 
  \hline
 $<$ 1950 & 24 & 0.43\% \\ 
  1950-1959 & 44 & 0.79\% \\ 
  1960-1969 & 634 & 11.39\% \\ 
  1970-1979 & 524 & 9.42\% \\ 
  1980-1989 & 780 & 14.02\% \\ 
  1990-1999 & 1615 & 29.03\% \\ 
  2000-2009 & 992 & 17.83\% \\ 
   $\geq$ 2010 & 950 & 17.09\% \\
   \hline 
     Total & 5564 & 100.00\% \\ 
   \hline
\end{tabular}
      \subcaption{Time Range of Data}\label{tab:timerange}
  \end{subfigure}   
  \begin{subfigure}[b]{.32\textwidth}
      \includegraphics[width=\textwidth]{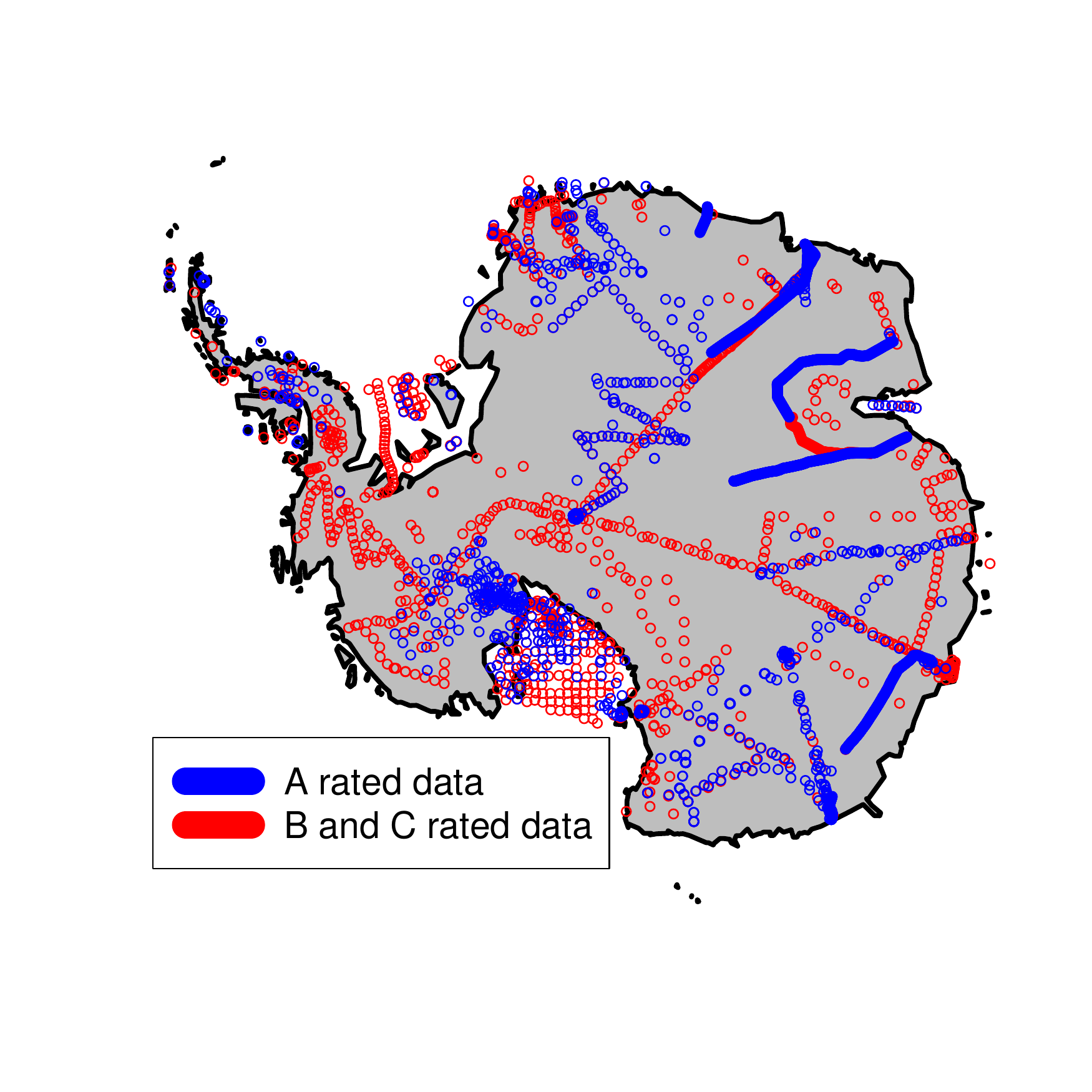}
            \vspace{-10mm}
      \subcaption{Data Location and Rating}
  \end{subfigure}
    \begin{subfigure}[b]{.32\textwidth}
      \includegraphics[width=\textwidth]{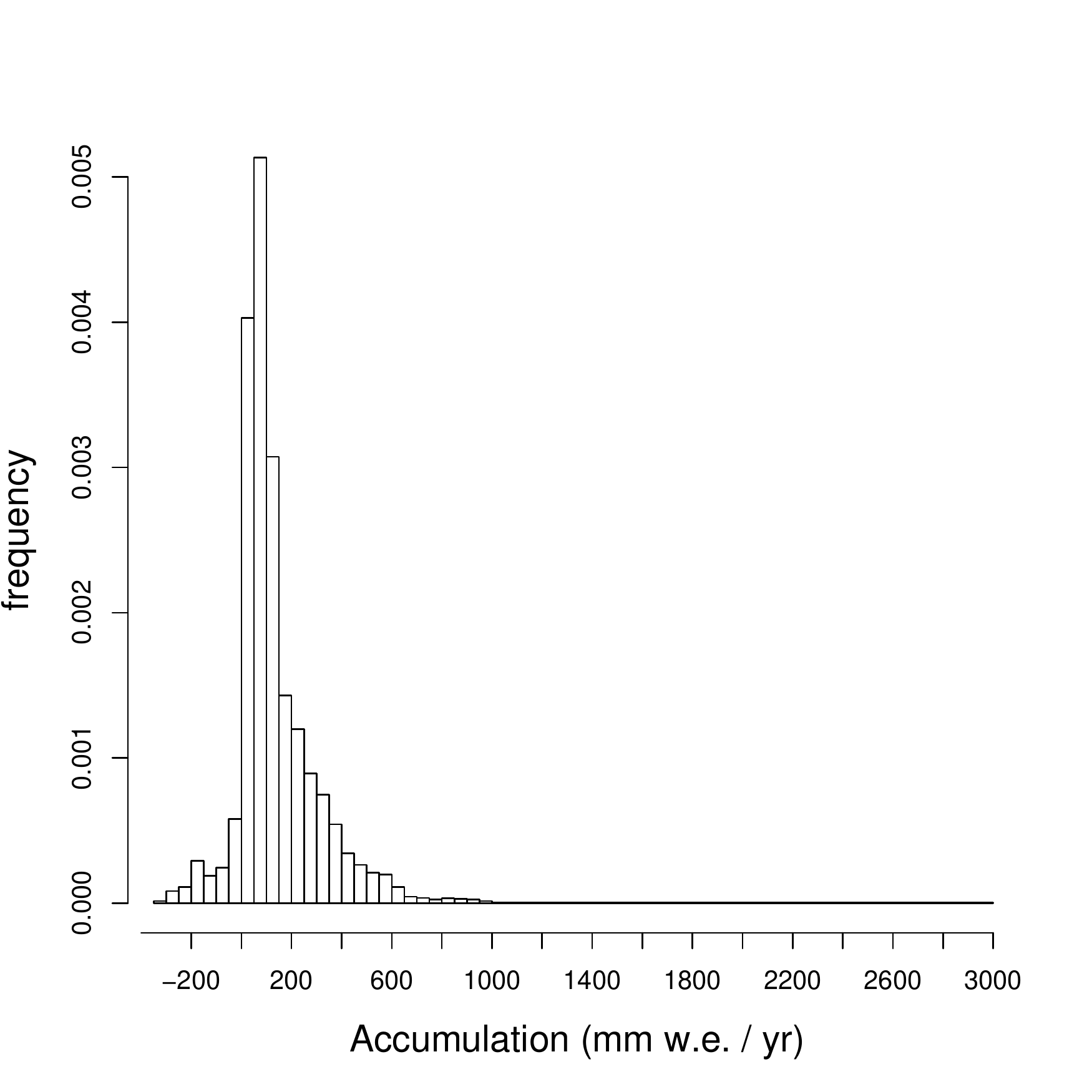}
      \vspace{-5mm}
      \subcaption{SMB histogram}
  \end{subfigure} 
  \caption[SMB Data Location]{(a) Summary of measurement time. Because many measurements are a summary of SMB over several years, we report the the final year of the measurement time range. (b) Aggregate dataset $N=5564$ mapped with reliability ratings: A or non-A-rated. (c) Histogram for SMB data. }\label{fig:alldat}
  \end{center}
\end{figure}

The dataset used in our analysis was aggregated by \citet{favier2013}, and consists of point-source time-averaged SMB measurements over the Antarctic ice sheet (i.e. measurements are the average rate of accumulation). The majority of the data comes post-1960. The temporal coverage of the data is given in Figure \ref{tab:timerange}. Using the reliability ratings suggested by \citet{magand2007}, \citet{favier2013} compiled 5564 SMB field measurements at $N_u = 5101$ unique locations from over 90 sources. For each data source, a reliability rating of ``A,'' ``B,'' or ``C'' was given depending upon the method and the duration of the measurement. We plot data locations and a data histogram in Figure \ref{fig:alldat}. 

For A-rated measurements, the minimum observed SMB is $-306$ mm w.e. $\textrm{yr}^{-1}$ and the maximum is $1665$ mm w.e. $\textrm{yr}^{-1}$. For all measurements, the minimum is $-317$ mm w.e. $\textrm{yr}^{-1}$ and the maximum is $2860$ mm w.e. $\textrm{yr}^{-1}$. Importantly, the available database includes labels for data only as A-rated or non-A-rated data, an issue we address in modeling decisions. While \citet{bromwich2011} argue for the cautious use of less reliable data, if we were to exclude all less reliable, non-A-rated data, then we would reduce the available data from 5564 to 3529 observations. More importantly, the A-rated-only dataset of 3529 measurements significantly reduces spatial coverage relative to the full dataset (see Figure \ref{fig:alldat}); however, even the ``A-rated only'' dataset contains more field measurements than those used in previous analyses \citep{vaughan1999,vandeberg2006,bromwich2004,lenaerts2012}. To cautiously use all available field data, as is emphasized by \citet{bromwich2011}, we pose a latent mixed model that accounts for additional uncertainty associated with less reliable measurement methods (B and C-rated data) as defined by \citet{magand2007}.
\vspace{-12mm}
\vspace{-5mm}
In the dataset, elevation, temperature, and distance to the coast values are missing for many observations. Because these are important quantities for projecting SMB \citep{vaughan1999}, we estimate these missing values using remote sensing data and climate reanalysis data. Specifically, we use the Radarsat Antarctic Mapping Project (RAMP) digital elevation model (DEM) with 200 m grid coarseness to estimate the elevation for coordinates missing elevation data \citep{liu2001}. The RAMP DEM is evaluated on two grids: the WGS84 ellipsoid and the OSU91A geoid, and the estimated elevations on these grids differ slightly. We impute missing values with the average of the elevation estimates on the WGS84 ellipsoid and the OSU91A geoid. To estimate temperature, we use the ERA Interim 2-m air temperature data averaged from 1979 to 2014 \citep{molteni1996}. Because the ERA Interim temperature grid is coarse, we use a weighted average of the eight nearest grid temperatures, weighted proportionally to inverse squared distance. 

Because using Euclidean distances can distort spatial relationships, leading to potentially inaccurate spatial interpolation and predictions \citep{banerjee2005}, we use great-circle (or spherical) distance, the shortest distance between two points on the surface of a sphere. The distance to coast is computed using the coast coordinates, where we consider the coast to be the edge of the ice sheet.  We utilize elevation, temperature, and distance to coast as covariates in our analysis because they have proven useful in predicting Antarctic SMB in other studies \citep[e.g.][]{vaughan1999}.

Because the data are right-skewed, asymmetric, and peaked (see Figure
\ref{fig:alldat}), it is likely that a Gaussian probability model could fail to capture some important data features. In addition, we computed the variance and mean within small binned regions and found them to be highly correlated ($r \approx 0.76$). That is, where higher SMB is observed, we observe higher spatial variability in SMB. Problematically, correlation between sample variance and mean can introduce bias in areas of spatial extremes \citep{christensen2011}. Because these data characteristics present potential modeling issues, we address them through data transformations. First proposed by \cite{box1964}, we consider the class of delta method-derived variance-stabilizing transformations (VST) laid out by, for example, \citet[chap. 3]{hocking2013}.

\vspace{-3mm}
\section{Covariance Modeling}\label{sec:covar}

We expect, geologically, that changes in spatial relationships (distance) are different as a function of elevation change. Thus, we do not assume that elevation and distance are separable, and we adapt classes of non-separable space-time covariance models to distance-elevation models. In this problem, the data are collected on a sphere (the earth) with a measurement of elevation. We represent our space as $D := \mathbb{S}^2 \times \mathbb{R}$, where $\mathbb{S}^2 := \{(x,y,z) \in \mathbb{R}^3 : \sqrt{x^2 + y^2 + z^2} = r \}$ and $r$ is the radius of the earth, as was done in \cite{gneiting2013}. Accordingly, we seek covariance functions suitable for $D$.

Ignoring the spherical nature of the data, we could build covariance functions on $\mathbb{R}^3 \times \mathbb{R}$. Let $\lVert \bh\rVert$ be the Euclidean distance between two arbitrary points in $\mathbb{R}^3$ and let $u$ be the difference between two elevations in $\mathbb{R}$. For this case, \cite{gneiting2002} developed a general class of non-separable covariance functions (the so-called Gneiting class). These covariance functions take the form 
\begin{equation}\label{eq:gneiting}
C(\bh ;u) = \frac{\sigma^2}{\bpsi( |u|^2 )^{d/2} } \bphi \left( \frac{ \lVert \bh \rVert^2}{\bpsi( |u|^2 )} \right),
\end{equation}
where $\bphi(s)$ is a completely monotone function for $x \geq 0 $ and $\bpsi(s)$ is a positive function with a completely monotone derivative (i.e. a positive-valued Bernstein function). \cite{zastavnyi2011} discuss the necessary and sufficient conditions for positive-definiteness of this class. The distance arguments can be inverted, which we call the inverted-Gneiting class, as is presented in \cite{porcu2016},
\begin{equation}\label{eq:invgneiting}
C(\bh ;u) = \frac{\sigma^2}{\bpsi(\lVert \bh \rVert^2 )^{d/2} } \bphi \left( \frac{ u^2}{\bpsi( \lVert \bh \rVert^2)} \right).
\end{equation}
Because Euclidean distance does not address the geometry of the earth it can lead to spatial distortion and poor model performance in some cases \citep{banerjee2005}. For this reason, these non-separable models may be inadequate for data on a sphere. 

\cite{gneiting2013} thoroughly discusses methods available to create spatial covariance functions using the great-circle distance. Following the notation of \cite{gneiting2013} and \cite{porcu2016}, we define $\bPsi_{d}$ to the class of positive-definite functions on $\mathbb{S}^d$, $\bPsi_{d,T}$ to be the positive-definite function class on $\mathbb{S}^d \times \mathbb{R}$, and $\bPsi_{\infty,T}$ to be the positive-definite function class on $\mathbb{S}^d \times \mathbb{R}$ for any $d$. For example, the Mat\'ern covariance function using the great-circle distance $\theta$ as an argument,
\begin{equation}\label{eq:circgneit}
C(\theta) = \sigma^2 \frac{1}{\Gamma(\nu) 2^{\nu-1}}  \left( \frac{\theta}{\rho} \right)^{\nu} K_{\nu} \left( \frac{\theta}{\rho} \right),
\end{equation}
is only a member of $\bPsi_\infty$ when $\nu \in (0,1/2]$ \citep[see Example 2 in ][]{gneiting2013}. Valid separable covariance models can be constructed by taking the product of valid covariance functions \citep[see, e.g.,][]{banerjee2014}. Therefore, the product of a covariance function from Equation \ref{eq:circgneit} and another valid covariance function is itself valid. 
\cite{porcu2016} extend the work of \cite{gneiting2013} on spatial covariance functions on spheres to create general non-separable space-time covariance classes valid using spherical distance. Theorem 1 of \cite{porcu2016} presents the inverted-Gneiting class for space-time problems. In particular, this class is of the form:
\begin{equation}\label{eq:circinvgneit}
C(\theta ;u) = \frac{\sigma^2}{\bpsi_{[0,\pi]}(\theta )^{1/2} } \bphi \left( \frac{ u^2}{\bpsi_{[0,\pi]}( \theta)} \right),
\end{equation}
where $\bphi(s)$ is, again, a completely monotone function for $x \geq 0 $, and $\bpsi_{[0,\pi]}(s)$ is the restriction of a positive-valued Bernstein function to $[0,\pi]$. This class of covariance functions belongs to $\bPsi_{\infty,T}$. 

By utilizing these classes, we assure that all covariance models considered are positive-definite. These classes provide a rich set of covariance functions on $\mathbb{S}^2 \times \mathbb{R}$, and we compare the predictive performance of these non-separable covariance functions to simpler covariance models, either separable or distance-only models.

\vspace{-3mm}
\section{Model Comparison}\label{sec:modcomp}
We consider a variety of spatial models for these data and compare their predictive performance on 100 randomly selected A-rated holdout datasets of size 1000. Following the recommendation of \cite{bromwich2011}, only A-rated data for model validation because they are more reliable than non-A-rated data \citep{magand2007}. 

Because prediction is our primary modeling goal, we present the following predictive measures as a means of model comparison across models: predictive root mean squared error (PRMSE), 90\% prediction interval coverage, and continuous rank probability score (CRPS) \citep{gneiting2007}, where 
\begin{equation} 
\text{CRPS}(F_i,y_i) = \int^\infty_{-\infty} (F_i(x) -  \bone(x \geq y_i) )^2 dx = \bE|Y_i - y_i | - \frac{1}{2}\bE | Y_i - Y_{i'} | ,
\end{equation}
where $F_i$ is the predictive CDF of $Y$ and $\bone(\cdot)$ is the indicator function.
Because we are utilizing MCMC to fit our model, we use posterior predictive samples for a Monte Carlo approximation of CRPS using an empirical CDF approximation \citep[see, e.g.,]{kruger2016},
\begin{equation} 
\text{CRPS}(\hat{F}^\text{ECDF}_i,y_i) = \frac{1}{m} \sum_{j=1}^m |Y_j - y_i| - \frac{1}{2m^2} \sum_{j=1}^m \sum_{k=1}^m | Y_j - Y_k| ,
\end{equation}
where $m$ is the number of MCMC samples used. We then average $\text{CRPS}(\hat{F}^\text{ECDF}_i,y_i)$ over all held-out data. PRMSE quantifies how well the model captures the mean, interval coverage measures how well the model quantifies uncertainty, and CRPS is a metric that considers how well the predictive distribution matches the data. In this paper, we use CRPS as the most important comparative metric as it accounts for the performance of the whole posterior predictive distribution, unlike PRMSE or prediction interval coverage which rely only on the predictive mean and quantiles, respectively. For our purposes, interval coverage signals model adequacy; therefore, models with 90\% interval coverage deviating greatly, say 10\% or more, from 90\% are not considered. 

These metrics are used to answer five modeling questions of particular interest: (1) Most basically, does a spatial model improve prediction relative to a non-spatial model? (2) Does including a latent mixed model improve prediction relative to a model that ignores measurement rating? In particular, if we constrain data obtained with less reliable methods \citep{magand2007} to have higher variance, does this improve prediction. Details about this formulation are presented in more detail in Equation \ref{eq:latvar}. (3) Which covariance models most effectively capture details of the data? In particular, we consider many covariance models: (i) a distance-only model, (ii) a separable covariance model using great-circle distance and elevation, and (iii) a variety of non-separable covariance models from the Gneiting and inverted-Gneiting classes. (4) Do variance stabilizing transformations improve predictive performance? (5) If we allow covariate effects to change over space through a multivariate spatial process, are predictions more accurate \citep{gelfand2010,datta2016a}? 

To answer these questions, we compare 34 models that differ with regards to covariance model, variance stabilization, GP specification, the inclusion of a latent mixed model, and whether spatially-varying covariate effects are used. In Section \ref{sec:nngp}, we discuss properties of nearest-neighbor Gaussian processes that are used to fit the models being compared.  We then compare competing models in Section \ref{sec:mod_spec_res}.

\vspace{-2mm}
\subsection{Nearest-Neighbor Gaussian Processes}\label{sec:nngp}

A Gaussian process (GP) model is a stochastic process, denoted GP$(m(\bs), C(\bs,\bs'))$, for which any finite collection of random variables from the process are jointly Gaussian and are fully specified by its mean function $m(\bs)$ and covariance function $C(\bs,\bs')$ \cite[see, e.g.,][]{banerjee2014}. The mean function defines the center of the process as function of the model space, and the covariance function governs the smoothness and uncertainty associated with the process. Gaussian processes are natural choices to model spatially varying phenomena because of the flexibility in specifying correlation between points in a compact neighborhood expressed by the covariance function \citep{stein1999,banerjee2014,cressie2015}.

Nearest-neighbor Gaussian processes (NNGP's) induce sparsity in the precision matrix by assuming conditional independence given reference sets \citep{datta2016a,datta2016b}. Suppose we begin with a parent GP over $\mathbb{R}^d$, then the GP is completely specified by its mean and cross-covariance function \citep{gelfand2010}. Then, the NNGP requires selecting a reference set $\mathcal{S} = \{\bs_1,\bs_2,...,\bs_k \}$ of $k$ distinct locations, where we impose an ordering on the $k$ locations. Then, we define neighborhood sets $N_\mathcal{S} = \{N(\bs_i); i = 1,...,k \}$ over the reference set with $N(\bs_i)$ consisting of the $m$ nearest-neighbors of $\bs_i$, selected from $\{\bs_1,\bs_2,...,\bs_{i-1} \}$. Note that if $i \leq m+1$, $N(\bs_i) = \{\bs_1,\bs_2,...,\bs_{i-1} \}$. Along with $\mathcal{S}$, $N_\mathcal{S}$ defines a directed acyclic graph (DAG). The joint distribution of $\bw_\mathcal{S}$, the Gaussian DAG, can be expressed as 
\begin{equation}
[ \bw_\mathcal{S} ] = \prod^k_{i=1} [\bw(\bs_i)|\bw_{N(s_i)}]
= \prod^k_{i=1} \mathcal{N}(\bw(\bs_i)| \bB_{\bs_i} \bw_{N(s_i)}, \bF_{\bs_i}), \nonumber
\end{equation}  
where $\mathcal{N}$ is the normal distribution, $\bB_{\bs_i} = C_{\bs_i,N(\bs_i)} C_{N(\bs_i)}^{-1}$, $\bF_{\bs_i} = C(\bs_i,\bs_i) - \bB_{\bs_i} C_{N(\bs_i),\bs_i}$, and $\bw_{N(\bs_i)}$ is the subset of $\bw_\mathcal{S}$ corresponding to neighbors $N(\bs_i)$ \citep{datta2016a}. \cite{datta2016a} extends this Gaussian DAG to a Gaussian process. Note that this GP formulation only requires us to store $k$ $m \times m$ distance matrices and requires considerably fewer floating point operations than the full GP model \citep[see][]{datta2016a}. Like any other GP model, the NNGP can be utilized hierarchically for spatial random effects. In this article, we use NNGP's as an alternative to the full GP specification. 

\vspace{-2mm}
\subsection{Model Specifications and Results}\label{sec:mod_spec_res}

As mentioned, the models compared in this section differ with respect to covariance model, variance stabilization, GP specification, and whether spatially-varying covariate effects are used. For covariance model comparisons, we consider great-circle distance $d$ using a Mat\'ern covariance function with $\nu \leq 1/2$; a separable covariance model $ C_1(d) C_2(u)$, where both $C_1$ and $C_2$ are Mat\'ern covariance functions and $\nu \leq 1/2$ for $C_1$; and several non-separable covariance functions. Specifically, we considered the following non-separable covariance models: (i) Equation (16) in \citet{gneiting2002} with $\nu = 1/2$ and $\nu = 3/2$ using chordal distance and (ii) variations on Equations (8)-(12) in \citet{porcu2016}. 

In this case, we use $u$, which normally refers to difference in time, to be elevation change. For this application, \citet{porcu2016}'s direct construction covariance models (Equations (10)-(12) from \citet{porcu2016}) were not well conditioned for short spherical distances. Ultimately, we found that a simplified version of Equation (8) in \cite{porcu2016} gave the best predictive performance in terms of CRPS compared to other non-separable covariance models. Therefore, all the presented results for non-separable covariance models refer to this covariance model:
\begin{equation}\label{eq:nonsep}
C(d,u) = \frac{\sigma^2}{\left\lbrace 1 + \left( \frac{d}{\rho_1} \right)^\alpha \right\rbrace^{\delta + \nu/2} } \exp\left[ - \frac{\left( \frac{u}{\rho_2} \right)}{\left\lbrace 1 + \left( \frac{d}{\rho_1} \right)^\alpha \right\rbrace^{\nu/2} } \right],
\end{equation} 
where $\rho_1,\rho_2 >0$ are range or scale parameters for spherical distance and elevation change, respectively, $\alpha \in (0,2]$, $\sigma^2 >0$, $\delta > 0$, and  $\nu \in[0,1]$ determines the degree of separability. If $\nu = 0$, then we have a separable covariance function with generalized Cauchy covariance for spherical distance and an exponential covariance for elevation change.

While several VST's were considered, the Box-Cox transformation was most effective stabilizing the variance-mean correlation, reducing sample correlation from 0.76 to 0.23. Therefore, we use the Box-Cox transformation as a comparison to models using non-transformed data. Because the data have several negative values, we add a constant ($306.001$ mm w.e. $\textrm{yr}^{-1}$) to make all values to positive, allowing us to perform the Box-Cox transformation. For simplicity, we use maximum likelihood estimation to estimate the transformation parameter $\lambda$ \citep[see][for details]{box1964}. 

Lastly, we consider multivariate NNGP's for spatially-varying coefficients \citep{gelfand2010} and compare their predictive performance to models only using spatially-varying intercepts. For the various spatial correlation functions discussed ($C(\bs,\bs')$), we used cross-covariance functions where $\text{Cov}(\bw(\bs),\bw(\bs')) = \bV C(\bs,\bs')$ for multivariate NNGP's, where $\bV \sim \text{Inverse Wishart}(\bI,p+1)$ \emph{a priori} and represents between-covariate covariance. To clarify, SMB is univariate; multivariate NNGP's refer to spatially-varying regression coefficients. 

For all NNGP models, we found that the NNGP predicted as well as full GP's with $m=10$ neighbors, and we did not observe predictive benefits beyond $m=10$ neighbors in terms of CRPS; however, we chose $m=20$ neighbors to be conservative. Additionally, we select the reference set to be unique data locations ($k=N_u$). All models take the form
\begin{equation}
Y_{ij} = \bx_{ij}^T \bbeta + \bz_{ij}^T \bw_i + \epsilon_{ij},
\end{equation}
where $i$ indexes location, $j$ indexes repeated measurements, $\bx_{ij}$ are covariates or interactions with fixed effects $\bbeta$, $\bz_{ij}$ are covariates with spatially-varying effects $\bw_i$, and $\epsilon_{ij}$ is random noise. Additional modeling details are presented in Section \ref{sec:mod} and model fitting details are given in Appendix \ref{sec:Gibbs}. All models are compared on the original scale of the data, requiring us to back-transform predictions from the Box-Cox space into the original scale of the data. The model comparison criteria are given in Table \ref{tab:modelcomp} for all models considered.

The first point to note, given the results in Table \ref{tab:modelcomp}, is that a spatial analysis of this data is justified. Furthermore, models that utilized spherical (or great-circle) distance have lower PRMSE and CRPS relative to those using Euclidean distance, even though this restricts the class of covariance functions that can be used. For most cases, the latent mixed model improves prediction relative to models that did not attempt to differentiate between B and C-rated measurements. Ultimately, we select model 18 from Table \ref{tab:modelcomp} because it had the lowest CRPS, adequate 90\% interval coverage, and low PRMSE. This model uses a non-separable covariance function using spherical distance, spatially-varying regression coefficients, and the Box-Cox transformation. We carry out the remainder of the analysis using this model.
\vspace{-3mm}
\section{Model Specification}\label{sec:mod}
We adopt a Gaussian likelihood because of the Box-Cox transformation to normality, where observations are conditionally independent given the modeled mean and random effects ($\bx_{ij}^T \bbeta + \bz_{ij}^T \bw_i $). This assumption is assessed in Section \ref{sec:res}. We construct a Gaussian random field through NNGP models to flexibly and accurately represent SMB as a function of the spatial arrangement of data collection sites. Moreover, NNGP models yield accurate predictions and rigorous uncertainty quantification.

Because our data include multiple measurements at some locations, our model is written for an arbitrary number of measurements at each site. We pose a covariance model for $\bw(\bs)$ using great-circle distance $d$ and elevation change $u$, which are functions of the spatial location $\bs =$(latitude $\phi$, longitude $\lambda$, elevation $E$). Our model takes the form
\begin{align}
Y_{ij} &= \bx_{ij}^T \bbeta + \bz_{ij}^T \bw_i + \epsilon(y^*_{ij}), \\
\bw(\bs) &\sim \text{NNGP}(0,C(\bs,\bs')), \nonumber \\
C(\bw(\bs),\bw(\bs'))&=\bV C(d,u).\nonumber
\end{align}
where $C(d,u)$ is defined in Equation \ref{eq:nonsep}, $i$ indexes location, $j$ indexes repeated measurements, $Y_{ij}$ is the centered and scaled Box-Cox transformed SMB at location $i$ and repetition $j$,
\begin{align*}
\bx_{ij} &= (\text{el}_i , \text{dc}_i , \text{temp}_i , \text{el}_i \times \text{dc}_i ,
 \text{el}_i \times \text{temp}_i , \text{dc}_i \times \text{temp}_i , \text{el}_i \times  \text{dc}_i \times \text{temp}_i)^T, \\
 \bz_{ij} &= (1,\text{el}_i , \text{dc}_i , \text{temp}_i )^T,
\end{align*}
where $\text{el}_i$ is elevation, $\text{dc}_i$ is distance to coast, $\text{temp}_i$ is 2-m air temperature, and each term is centered and scaled so that it has a mean of zero and a standard deviation of one. We take the product, indicated by $\times$, of these centered and scaled covariates to give interaction effects. Covariate effects are denoted by $\bbeta$, $w_i$ are spatially-varying coefficients, $\epsilon(y^*_{ij})$ is Gaussian noise indexed by a latent variable $y^*_{ij}$. 

Because there is no distinction between B-rated and C-rated measurements within the available database, we introduce $y^*_{ij}$ to distinguish B and C-rated data, each with unique distributional assumptions. To account for error associated with different measurement techniques, we include multiple nugget effects (error terms) $\tau^2_A$, $\tau^2_B$, and $\tau^2_C$ corresponding to the measurement ratings given by \citet{magand2007}. Specifically,
\begin{equation}\label{eq:latvar}
\begin{aligned}[c]
\epsilon_{ij} |y^*_{ij} = A  &\sim N(0,\tau^2_A), \\
\epsilon_{ij} |y^*_{ij} = B  &\sim N(0,\tau^2_B), \\
\epsilon_{ij} |y^*_{ij} = C  &\sim N(0,\tau^2_C), 
\end{aligned}
\hspace{10mm}
\begin{aligned}[c]
&\text{Pr}(y^*_{ij} = B | y^*_{ij} \neq A ) = \theta , \\
&\text{Pr}(y^*_{ij} = C | y^*_{ij} \neq A ) = 1-\theta, \\
\end{aligned}
\end{equation}
where we constrain Pr$(\tau^2_a<\tau^2_b<\tau^2_c)=1$. The latent variable $y^*_{ij}$ attempts to capture the differences in reliability between B-rated and C-rated measurements since the database does not distinguish between B and C-rated data. We also considered the need of additive and multiplicative errors for B and C-rated data in preliminary models; however, these effects did not improve predictive performance. 

This model extends the non-separable space-time covariance models to non-separable spatial quantities (elevation and spherical distance). We also discuss the NNGP for use with an arbitrary number of measurements at the same location. Additionally, our latent variable formulation allows data of varying reliability levels to be incorporated. This model is utilized for model-based design in Section \ref{sec:prop}.
  
  \vspace{-2mm}
\subsection{Priors, Model Fitting, and Prediction} 

We use the following prior distributions on hyperparameters $\tau^2_A$, $\tau^2_B$, $\tau^2_C$, $\theta$, $\bbeta$, $\bV$, $\rho_1$, $\rho_2$, $\nu$, $\alpha$, and $\delta$:
\begin{equation}
\begin{aligned}[c]
  \tau_A^2 &\sim \text{IG}(20,6), \\
  \tau_B^2 &\sim \text{IG}(20,8), \\
  \tau_C^2 &\sim \text{IG}(20,10), \\
  \theta &\sim \text{Beta}(1,1),
    \end{aligned}
    \qquad
    \begin{aligned}[c]   
  \bbeta &\sim \mathcal{N}( \bzero, \bI ), \\
      \rho_1 &\sim \text{Gamma}(2,20), \\
       \rho_2 &\sim \text{Gamma}(1,10), \\
\bV &\sim \text{IW}(\bI,p+1),
  \end{aligned}
      \qquad
  \begin{aligned}[c]
  \alpha &\sim \text{Unif}[0,2], \\
  \nu &\sim \text{Unif}[0,1], \\
  \delta &\sim \text{Gamma}(1,1), \\
    \end{aligned}
  \end{equation}
where the Gamma and Inverse Gamma distributions use the shape-rate parameterization. In the presence of spatial random effects, \cite{hodges2010} demonstrate that the behavior of fixed effects is unpredictable. Due to centering and scaling our data, inclusion of the spatial random effect $w_i$, and the unknown role of covariates above and beyond the spatial random effects, we choose $\bm_\beta = \bzero$. To select the prior distribution for $\rho_1$ and $\rho_2$, we consider several things. \citet{vandeberg2006}, who utilize a weighted average approach, suggested smoothing model output within 193 km because this yielded predictions most correlated with observations. On the other hand, \citet{arthern2006} suggest that there is no range where a semivariogram reaches its sill (i.e. the distance where points are no longer correlated); however, they fit the semivariogram with a line and consequently could not estimate a finite range. Also, locations on the Antarctic ice sheet up to 570 km from the nearest A-rated data point. For these reasons, we have selected a diffuse prior distribution with \emph{a priori} mode near $ 0.10$ (range $\approx$ 570 km).
Thus, our prior structure is flexible and will allow the estimated spatial process to converge to the covariance process of Antarctic SMB. Then, we choose prior distributions on $\sigma^2$, $\tau^2_A$, $\tau^2_B$, and $\tau^2_C$ to be relatively diffuse but match the scale of our centered and scaled outcomes. Lastly, we select $a_\theta=1$ and $b_\theta=1$ because we do not know whether to favor $B$ or $C$ rated data with any certainty.

We sample from the posterior distribution via Markov chain Monte Carlo (MCMC) using a Gibbs sampler for all parameters except for covariance parameters $\rho_1$, $\rho_2$, $\alpha$, $\delta$, and $\nu$ for which we use the Metropolis-Hasting algorithm to sample. Full conditional distributions are provided in Appendix \ref{sec:Gibbs}. Using Integrated nested Laplace approximation (INLA) is an alternative model fitting approach; however, MCMC samples from the posterior distribution are straightforward to utilize for prediction.

We predict SMB, which we denote as $\by_g$, on a stereographically uniform grid over the Antarctic ice sheets to compute integrals of interest, (e.g. average and net SMB) by drawing from the posterior predictive distribution,
\begin{equation}
f(\by_g | \by) = \int_\theta  f ( \by_g | \eta,\by,\bX) \, \pi( \eta | \by,\bX) \, \textrm{d}\eta,
\end{equation}
where $\eta$ represents all model parameters using composition sampling \citep[see, e.g.,][]{gelman2014}. Importantly, prediction requires selection of $m$-nearest-neighbors from the reference set $\mathcal{S}$ for each grid location. Then, spatial random effects at grid locations follow a conditional normal distribution, where conditioning is limited to each location's neighbors. For any location $\bs$,
\begin{equation}
\bw(\bs) \vert \bw_{\mathcal{N}(\bs)} \sim \mathcal{N}\left( C_{\bs,N(\bs)} C_{N(\bs)}^{-1} \bw_{N(\bs)} , \\
C(\bs,\bs) - C_{\bs,N(\bs)} C_{N(\bs)}^{-1} C_{\bs,N(\bs)}^T \right), \label{eq:cond}
\end{equation}
where $\bw_{N(\bs)}$ are the observed random effects at the neighbors of $\bs$. To estimate mean SMB over Antarctic ice sheet $\mathcal{D}$
\begin{equation}
\overline{SMB} = \frac{1}{|A|}\int_{\mathcal{D}} \by_g f(\by_g \mid \by)\textrm{d} A_{y_g} \text{d}\by_g ,
\end{equation}
where $\textrm{d} A_{y_g}$ accounts for the area distortion due to the stereographic projection and $|A|$ is the area of the Antarctic ice sheets. The integral over the posterior distribution accounts for sampling variability due to the measurements of SMB, as well as model parameter uncertainty.
Using predicted SMB at grid values, we estimate the mean SMB and a 95\% credible interval for mean SMB.
We also estimate net SMB by integrating over predicted SMB again with respect to area $A$
\begin{equation}
\text{net SMB}=\int_{\mathcal{D}} \by_g f(\by_{\textrm{new}} \mid \by) \textrm{d} A_{y_g} \textrm{d} \by_g.
\end{equation} 
We quantify our uncertainty using a 95\% credible interval about net SMB and compare our results for the SMB map and net SMB estimates to previous results.

\vspace{-2mm}
\subsection{Extensions to Space-Time and Computer Model Emulation}\label{sec:extension}

While not implemented within this article, we briefly discuss extensions of this model for spatiotemporal data and for models that synthesize field measurements and output from deterministic mathematical or computer models (e.g. partial differential equation models). In the case of time-series data, one could imagine that the data could be associated with discrete or continuous time. For discrete time, neighbors could still be selected using great-circle distance. The NNGP model is amenable to the dynamic linear model framework proposed by \cite{west1997}, and the extension is discussed briefly in \cite{datta2016a}. In the case of continuous time, the modeling is like the current setting except that the selection of neighbors and the covariance model would incorporate a temporal component (see \cite{banerjee2014} for some discussion on space-time covariance). Extensions of the NNGP into spatiotemporal applications are discussed in greater detail in \cite{datta2016c}; however, this article focuses primarily on measurements that are taken at equal time intervals.

For a combination of field measurements $y_f$ and computer output $y_c$, we could pose a synthesis model of the form
\begin{equation}
y_f(\bs) =  a(\bs) + b(\bs) y_c(\bs) + \epsilon(\bs) \\
\end{equation}
where the additive and multiplicative discrepancy terms account for systematic differences between $y_c(\bs)$ and $y_f(\bs)$. For this model, we would assume that $a(\bs)$ follows an NNGP and $b(\bs)$ follows a log-NNGP. In this way, we account for biases in the computer model, enabling computer output to be used in addition to field measurements.

\vspace{-3mm}
\section{New Field Measurement Proposal Method}\label{sec:prop}
For any potential site of interest $\bs_g$, where $g$ indexes potential design locations, posterior predictive samples can be used to calculate integrated mean square error (IMSE). Explicitly,
\begin{align}
\textrm{IMSE}(\bs_g) &= \int_{\mathcal{D}} \textrm{MSE}[f(y(\bs) | \by , y_g)] \textrm{d} \bs \ \\
&= \int_{\mathcal{D}} E( f[y(\bs) | \by , y_g] - E \{ f[y(\bs)| \by , y_g] \} )^2  \textrm{d}\bs, \nonumber
\vspace{-6mm}
\end{align}
where $\bs \in \mathcal{D}$, $\mathcal{D}$ is the entire design space (the Antarctic ice sheet), and $y_g$ is the predicted SMB at $\bs_g$. In practice, this integral is computed either numerically or using Monte Carlo methods with samples from the posterior distribution. Integrated mean square error is calculated at potential design locations and quantifies the effect that a new measurement would have on uncertainty over the entire design space. Then, the design location with the lowest IMSE is proposed. In this way, the selection criterion is similar to expected improvement in the Bayesian optimization literature \citep[see, e.g.,][]{snoek2012}. Unlike many pre-data designs, we compute IMSE in the presence of previously sampled measurements, as was done by \cite{ranjan2011} applied to computer experiments. While \citet{ranjan2011} compared batches of proposed values, we carry out this procedure sequentially so that site proposals can be ranked in order of priority. 

This ranking is useful for glaciologists interested in exploring areas that will most improve future inference about SMB. Because observations are proposed but not taken sequentially, we treat posterior predictions at previously proposed locations as data. For example, when proposing a third site, we condition on the posterior predictive distributions of the first and second proposed locations as though they are data. By considering the entire posterior predictive distribution of proposals, we account for uncertainty in predictions at unobserved locations. Let $\bY_\text{dat}$ denote the vector of observed data at locations $\mathcal{S}_{dat}$. Algorithm \ref{alg:IMSE} presents a fully Bayesian iterative IMSE site selection and yields a set of potential measurement sites $\mathcal{S}_\text{prop}$. One could simplify this algorithm, using only posterior point-estimates instead of the entire posterior distribution. This would eliminate the innermost for-loop; however, this would not completely account for model uncertainty. This algorithm is presented in the Supplemental Material.

While IMSE can be prohibitively expensive computationally in big data settings under a full GP specification, the NNGP enables scalability of IMSE computation. When the mean and variance are correlated, as they are in this problem, one potential disadvantage of using IMSE as a design criterion is its tendency to propose new measurements in high variance areas instead of exploring the space of interest. Thus, IMSE can sacrifice exploration in favor of optimization.

We sequentially propose 25 measurement locations. These proposed measurement sites will provide Antarctic research locations that both fill the Antarctic ice sheet and have high uncertainty in SMB. Thus, we identify potential measurement locations that will reduce future uncertainty in total Antarctic SMB estimates. If desirable, this design scheme can be carried out regionally to identify areas of interest on smaller scales. It is interesting to consider including measurement type in the site selection method. In our problem, travel and thus data acquisition are so expensive that we plan to only propose and take A-rated measurements. However, in many problems, this is not the case (e.g. pollution monitoring). One could incorporate a utility function that takes both measurement type and location as arguments. This function would be subject to constraints that account for the cost specific to each measurement type.

\vspace{-2mm}
\begin{algorithm}[H]
\caption{Fully Bayesian Sequential Site Selection using IMSE}\label{alg:IMSE}
\noindent \textbf{Input}: $m$ samples from the posterior distribution for all model parameters, $\mathcal{S}_\text{prop}$, $\mathcal{S}_{dat}$, $\mathcal{S}_\text{grid}$ with the associated area $A_k$ which each ${\bs_g}_k \in \mathcal{S}_\text{grid}$ represents, and number of desired measurement sites $n_s$ \\
 
\noindent \textbf{Output}: Proposed sites $\mathcal{S}^*$
\begin{algorithmic}[1]
   \State \textbf{Initialize:} $n_\text{iter} = 0$, $n_p = \text{Size}(\mathcal{S}_\text{prop})$, and $\mathcal{S}^* = \emptyset$
\While {$n_\text{iter} < n_s$}
   \For{$i = 1$ to $n_p$}
      \State $\mathcal{S}_\text{cand} = \mathcal{S}_{dat} \cup \mathcal{S}^* \cup \bs_{p_i} $ for $\bs_{p_i} \in \mathcal{S}_\text{prop}$. This is the new reference set.
      \State Recompute Neighbors of $\mathcal{S}_\text{grid}$ to $\mathcal{S}_\text{cand}$
      \For{$j = 1$ to $m$}
         \State Sample ${\bY_g^{(\lambda)}}_j$ from $p(\bY(\mathcal{S}_\text{grid}) | \bY(\mathcal{S}_\text{cand}))$ using the $j^\text{th}$ posterior sample.
         \State Transform ${\bY_g^{(\lambda)}}$ to the original scale ${\bY_g}$
      \EndFor
      \State Compute $IMSE(\bs_{p_i}) \approx \sum_{k=1}^{n_g} A_k \text{Var}({\bY_g}_k)$
   \EndFor
    \State Define $\bs_\text{new} = \{ \bs_{p_i} : \overset{\min}{\bs_{p_i}} IMSE(\bs_{p_i}) \}$ 
    \State Take $\mathcal{S}^* = \mathcal{S}^* \cup \bs_\text{new}$  
   \State $n_\text{iter} \gets n_\text{iter} + 1$ 
\EndWhile

\Return $\mathcal{S}^*$
\end{algorithmic}
\end{algorithm}
\vspace{-3mm}

\vspace{-3mm}
\section{Results}\label{sec:res}
The results that we present are based on 50,000 posterior draws after a burn-in of 10,000 iterations.
To assure that these parameters have converged to stationary distributions and are mixing well, we utilize Geweke convergence diagnostics \citep{geweke1991} and Heidelberger-Welch (HW) diagnostics for posterior stationarity \citep{heidelberger1981,heidelberger1983}.
For the Geweke diagnostic, we calculate a $z$-score comparing the first 10\% to the last 50\% of the Markov chain and reject posterior stationarity if any $|z| >1.96$ for any parameter. Using HW diagnostics, we calculate the Cram\'er-von-Mises statistics and reject posterior stationarity if the computed p-value is less than $0.05$. For diagnostic results, see Table \ref{tab:parameter}. Because we observe $z$-scores less than $1.96$ for the Geweke diagnostic, no p-values below 0.05 for HW stationarity tests, we are satisfied with the mixing and convergence of this Markov chain. The Box-Cox transformation parameter was estimated to be $\hat{\lambda} = 0.347$ using maximum likelihood estimation. Posterior summaries are given in Table \ref{tab:parameter}. Note that elevation and temperature are positively related with SMB through $\beta_1 \text{ and } \beta_3$; however, their two-way interactions have a tempering effect on SMB ($\beta_4 \text{ and } \beta_6$). Similarly, when all covariates are high (or low) together, they appear to interact to increase (or decrease) SMB, on average (see $\beta_7$). Because elevation change and distance are rarely, if ever, zero while the other is non-zero, range parameters have limited interpretability with respect to effective range. 

\begin{figure}[H]
\begin{center}
\includegraphics[width = 0.35 \textwidth]{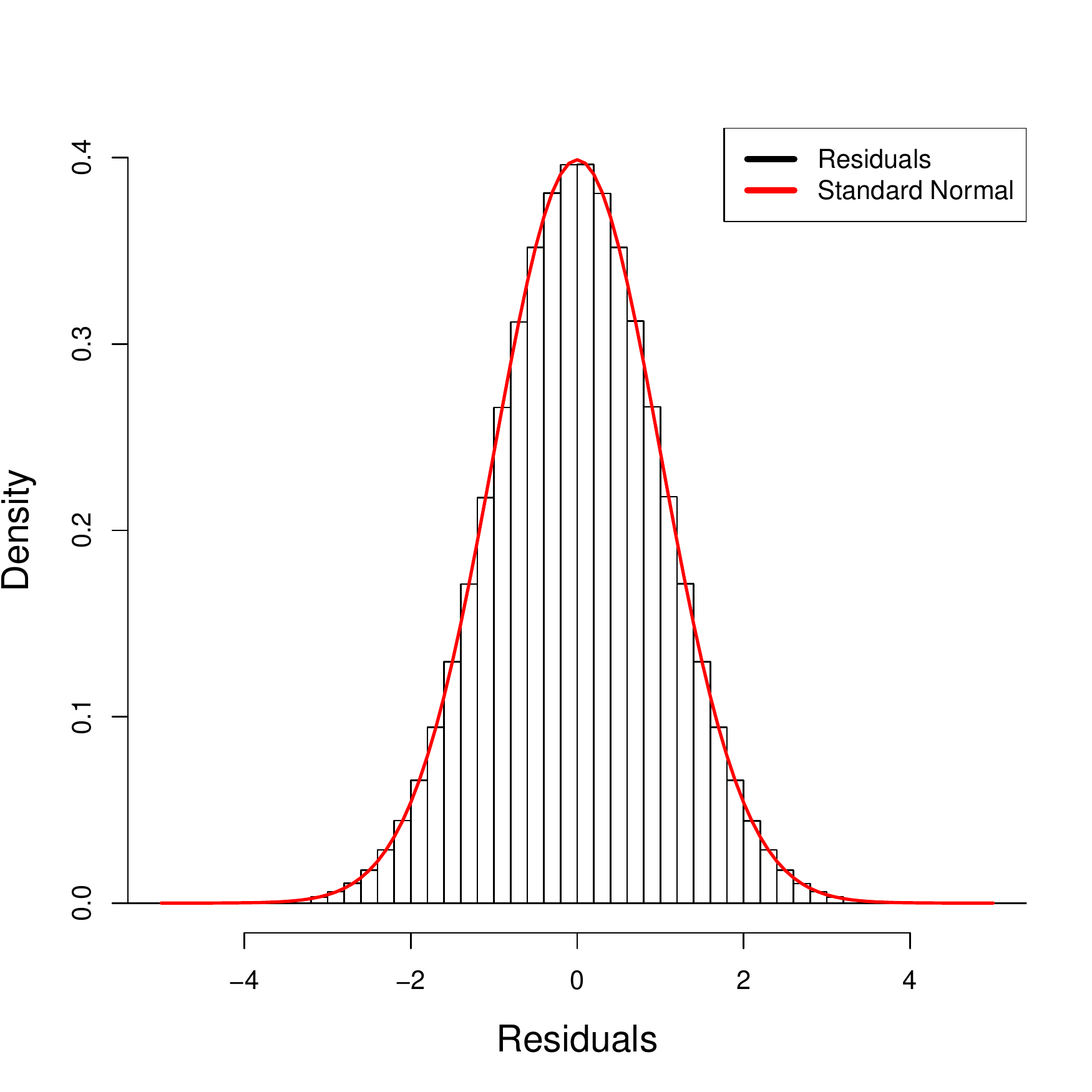}
\includegraphics[width = 0.35 \textwidth]{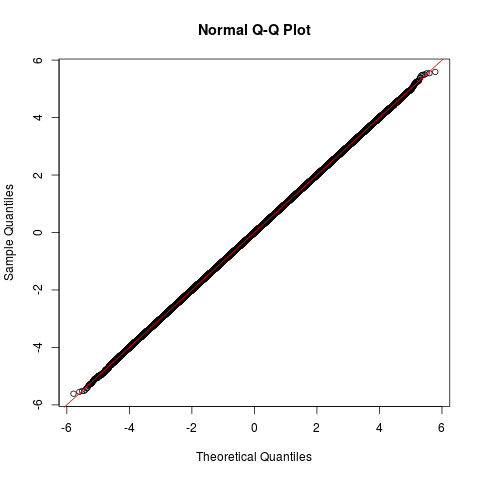}
\end{center}
\vspace{-5mm}
\caption{(Left) Scaled residual histogram for all posterior predictive samples. (Right) Normal qq-plot of all posterior predictive samples. }\label{fig:resid}
\end{figure}

Perhaps the most important model assumption to verify is conditional normality. Because the normal assumption is made after the Box-Cox transformation, we plot scaled residuals for the Box-Cox transformed data (see Figure \ref{fig:resid}). Given these residual plots, we are satisfied that the normal assumption is justified even though there are minor deviations from normality in the tails of the data. These residuals show that our model residuals has slightly lighter tails than the theoretical tails of the normal distribution. This aligns with our model having 93\% coverage for a 90\% prediction interval (See Table \ref{tab:modelcomp}).

\begin{figure}[H]
  \begin{center}
    \includegraphics[width=.48\textwidth,height=.4\textwidth]{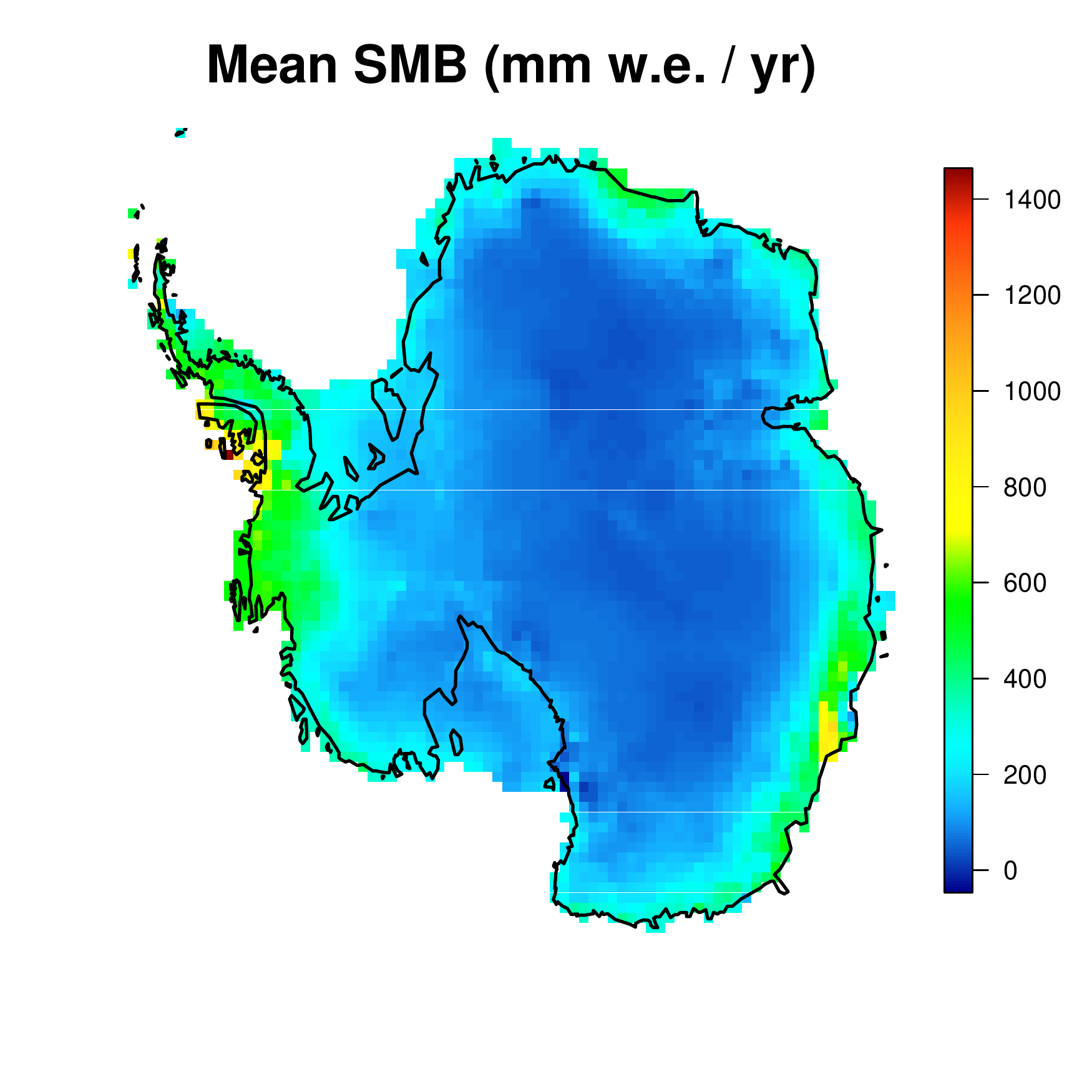}
    \includegraphics[width=.48\textwidth,height=.4\textwidth]{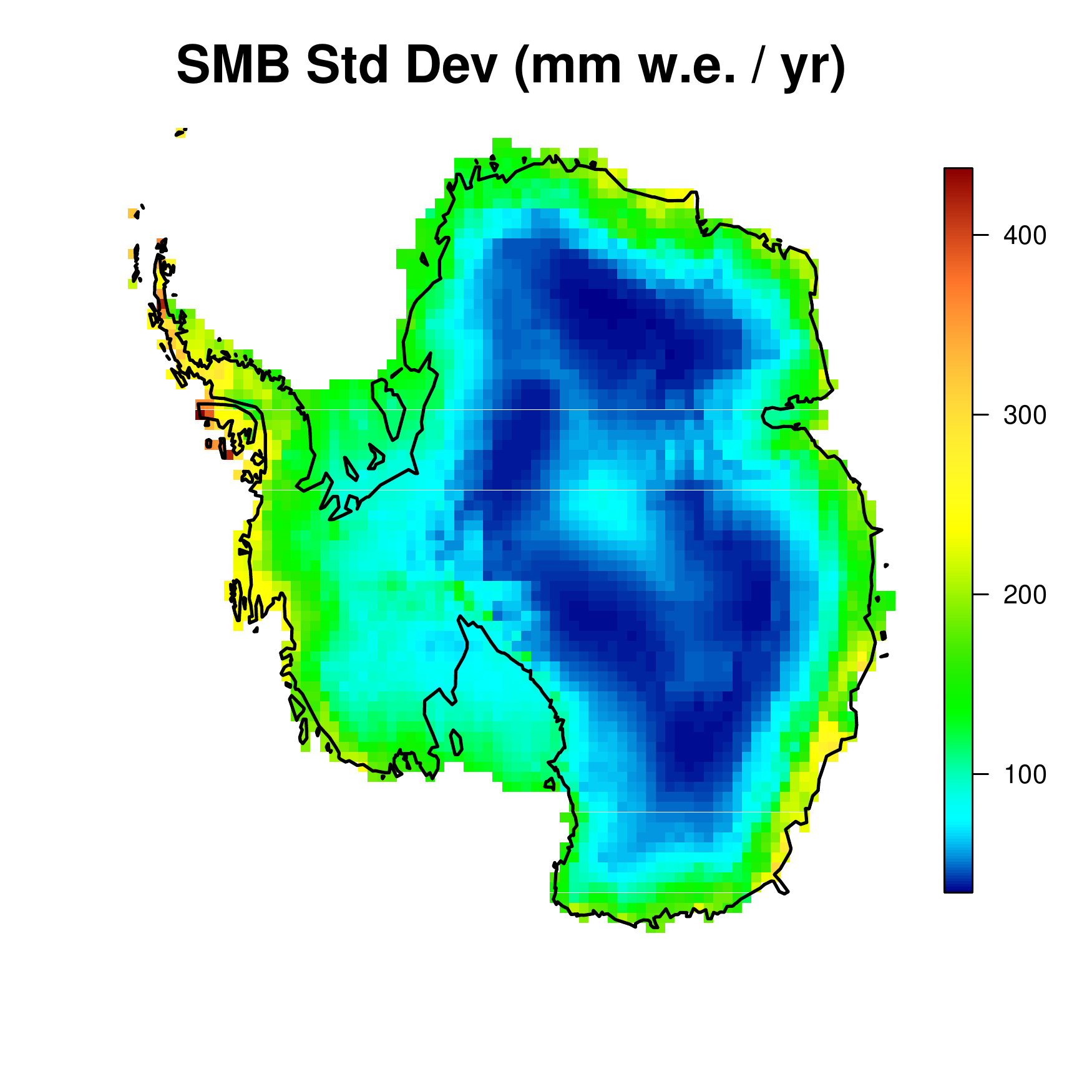} \\
  \end{center}
    \vspace{-10mm}
     \caption[SMB and prediction error surfaces]{SMB mean and prediction uncertainty maps (posterior predictive standard deviation) are plotted in units of mm w.e yr$^{-1}$. Note that high values in SMB and uncertainty are in red while low values are in blue. }\label{fig:surface1}
\end{figure}

\begin{figure}[H]
  \begin{center}
     \includegraphics[width=.48\textwidth,height=.4\textwidth]{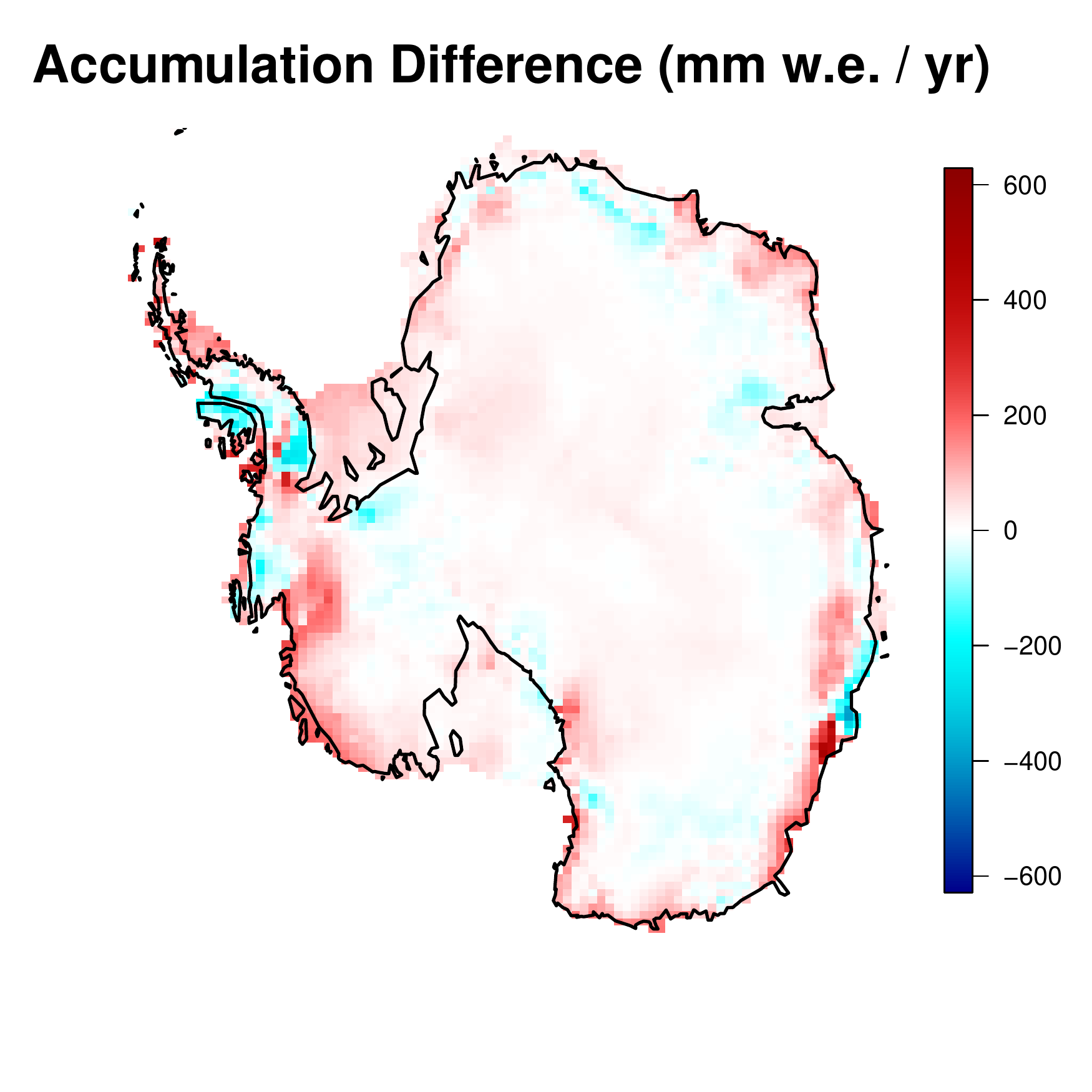}
    \includegraphics[width=.48\textwidth,height=.4\textwidth]{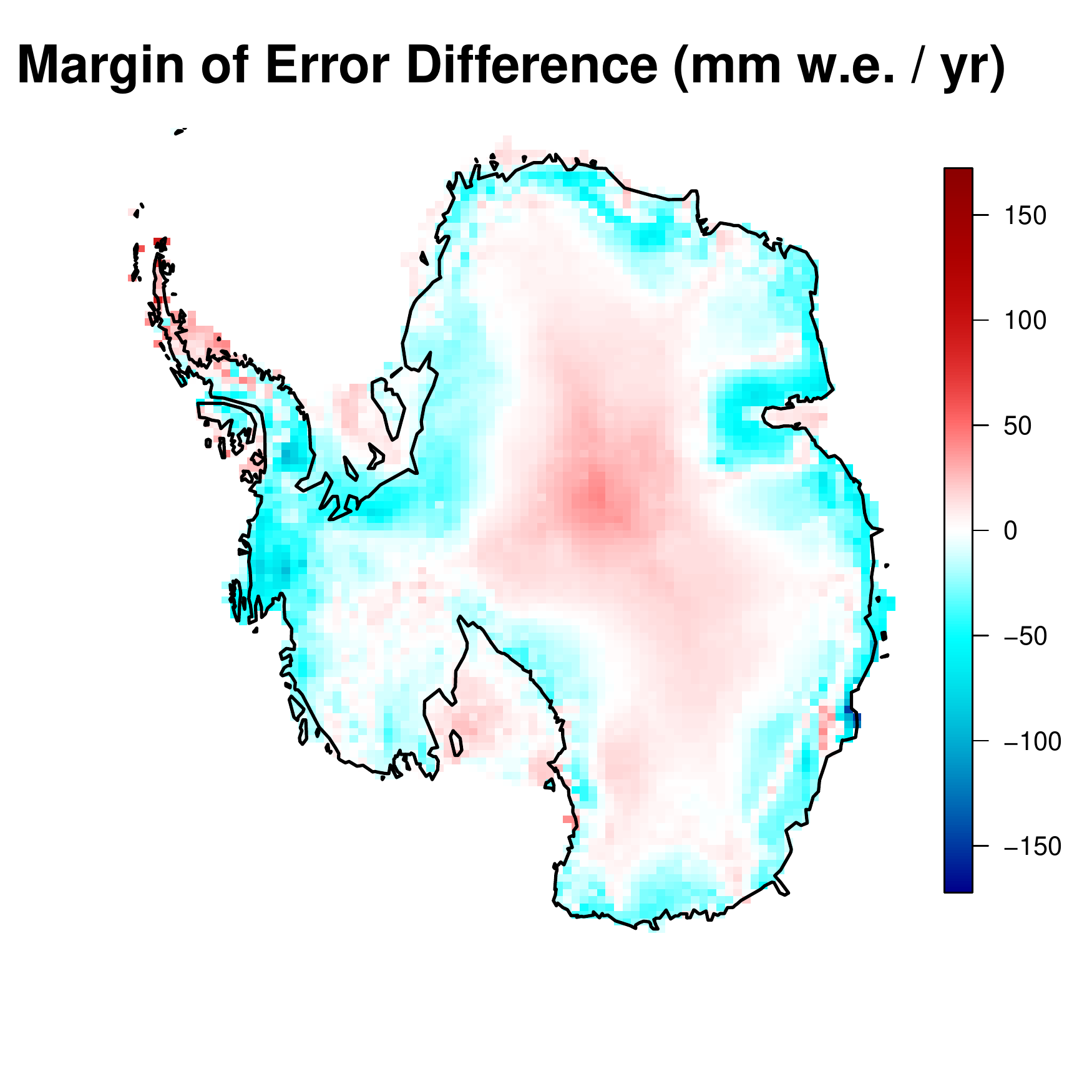} 
  \end{center}
    \vspace{-10mm}
     \caption[Difference  surfaces]{Differences in mean and prediction uncertainty (standard deviation) in SMB. These maps show the differences between predictions given by the model using all data compared to the equivalent model using only A-rated data.}\label{fig:surface2}
\end{figure}

\begin{figure}[ht]
  \begin{center} 
  \vspace{-20mm}     
  \includegraphics[width=\textwidth]{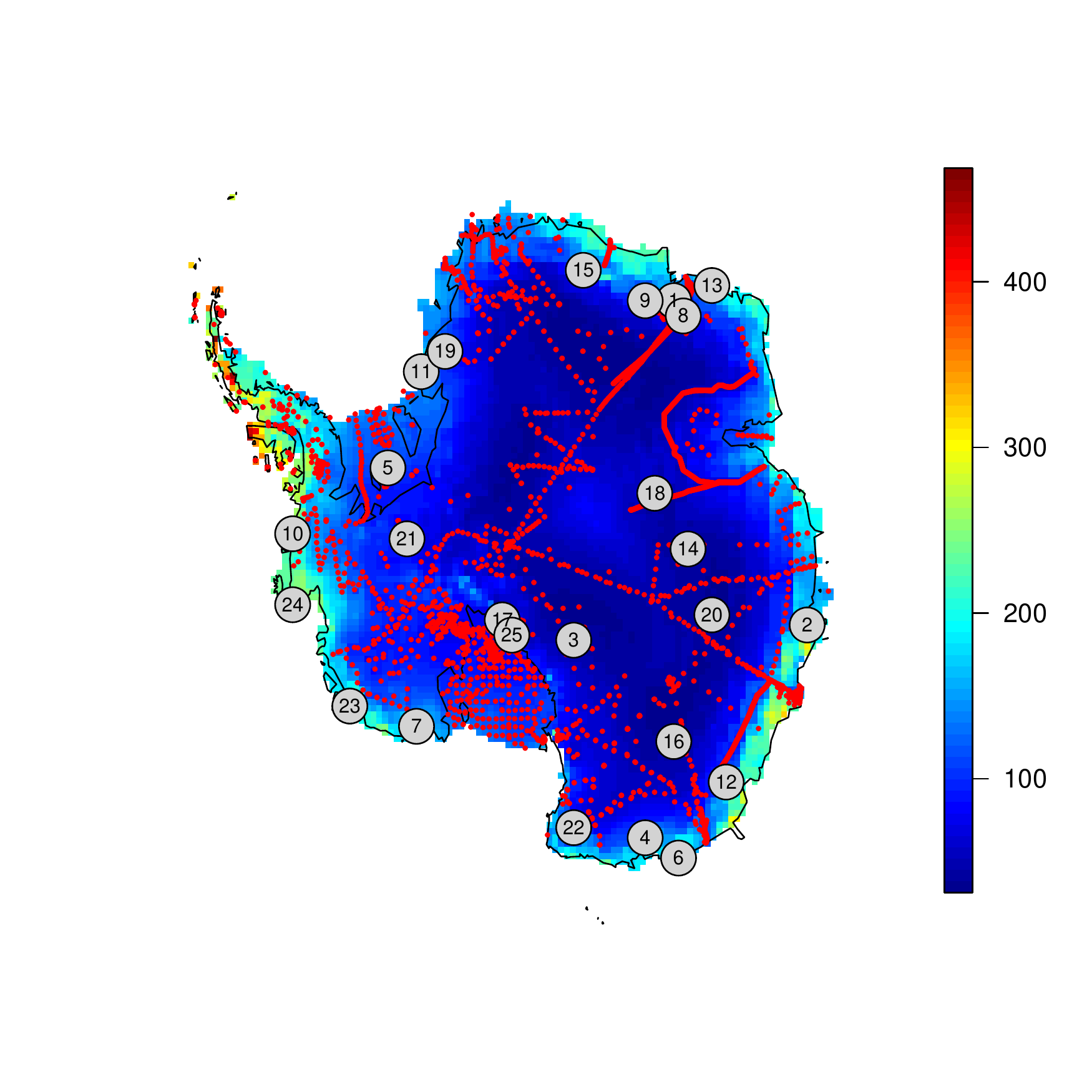}  
   \vspace{-30mm}
  \end{center}  
     \caption[SMB and prediction error surfaces]{Map of 25 new field proposal locations denoted by the numbered gray circles. The plotted number represents the order of measurement priority based on the iterative IMSE criterion presented. See Table 1 in Supplementary Material for coordinates. The proposal locations are plotted on top of the uncertainty map.}\label{fig:proploc}
\end{figure}

Using all $50,000$ post-burn-in draws from the posterior distribution of $\eta|\by$, we predict SMB to estimate net SMB, average SMB, and to create SMB maps. This requires back-transforming predictions from the Box-Cox space into the original scale of the data. Estimates for SMB are in Table \ref{tab:SMB} and Figure \ref{fig:surface1}. Using the mean of each grid prediction, we render an SMB heat map over the Antarctic ice sheet. Using posterior predictive standard deviation, we map prediction uncertainty spatially to identify regions of high model uncertainty. Both high SMB and high uncertainty in red and low SMB and low uncertainty in blue (see Figure \ref{fig:surface1}). As expected we see the highest SMB and uncertainty in western coastal regions. In general, the SMB map is similar to others that have been rendered \citep{vaughan1999,vandeberg2006,arthern2006, bromwich2004,lenaerts2012}.
Additionally, we plot the difference between the all-data and A-rated data models in Figure \ref{fig:surface2} to illustrate the changes in our estimates obtained by including non-A-rated data. Most significantly, the model using non-A-rated data gives higher coastal predictions in most regions relative to the A-rated-only model. 
\vspace{-4mm}
\vspace{-8mm}
\vspace{-4mm}
\vspace{-3mm} 
We use the fully Bayesian IMSE selection procedure presented in Section \ref{sec:prop}. The proposed measurements and their locations are shown in Figure \ref{fig:proploc}. Note that the design proposals are spread out but concentrated in areas of high SMB and consequently high uncertainty, especially coastal areas. For this reason, it is important that future field research focuses its efforts on studying and measuring these coastal regions that are poorly understood at this time, an argument shared by \citet{thomas2017}. Because this design scheme is meant to minimize integrated uncertainty in SMB, it will propose locations in high SMB and, thus, high uncertainty areas; however, we do not argue that studying lower SMB regions is not fruitful. For example, if determining whether SMB is positive or negative is the primary goal, then low SMB regions would be of particular interest. Another potentially important criteria to consider would be regions demonstrating significant temporal changes or variability, which is a topic of current research \citep[see, e.g.,][]{thomas2017}.
  \vspace{-5mm}
   
\section{Discussion and Conclusions}\label{sec:disc}
This analysis has two primary advantages compared to previous approaches characterizing SMB. First, its rigorous uncertainty quantification that allows us to identify regions where SMB estimation is most uncertain (see Figure \ref{fig:surface1}). Second, our analysis is the first to include all available data ($N=5564$) while accounting for measurement reliability and repeated measurements. Because \citet{vaughan1999} used about 1800 data points, \citet{arthern2006} made use of 540 data points, \citet{vandeberg2006} utilized 2032 data points, and \citet{lenaerts2012} used 750 data points, our model has utilized significantly more data than any other analysis. In addition (and in contrast) to previous work, we have modeled the spatial correlation for the great-circle distance and elevation change to account for spatial similarities and differences in SMB over the Antarctic ice sheets. Using both the great-circle distance and elevation change gives our model better predictive accuracy than previous SMB models \citep[see, e.g.,][]{arthern2006}. For these reasons, we argue that our estimates of SMB and associated uncertainty are more accurate than previous models. 

By comparing the uncertainty and SMB maps for the all data and A-rated data models, we identify advantages of including non-A-rated data. We can identify more SMB peaks, especially in coastal areas, by including all available data. On the whole, we see less uncertainty in our predictions when we include all available data, especially in regions lacking A-rated data; however, we observe increased uncertainty in regions where neighboring areas are rich in A-rated data and in some coastal areas due to the inclusion of non-A-rated data (see Figure \ref{fig:surface2}). Using our model, we can link areas of high prediction uncertainty with areas lacking data (Figure \ref{fig:surface2}) or exhibiting high climate volatility. Furthermore, our uncertainty quantification enables us to propose new field measurements designed to minimize integrated prediction error (Figure \ref{fig:proploc}). These proposed measurements provide valuable direction about which Antarctic regions could be studied in the future by climate scientists. Intuitively, our proposed measurements are in areas of high SMB and at boundary locations (i.e. coastal regions), areas we would expect high uncertainty. Like previous SMB estimation models, our model enables us to render SMB maps (see Figure \ref{fig:surface1}) that display regions of high SMB. These maps are vital to glaciologists for identifying or proposing climate drivers causing regional variability in SMB across Antarctica and the net mass balance of the Antarctic ice sheet.

\begin{figure}[H]  
  \begin{center}
   \begin{subfigure}[b]{.4\textwidth}
      \includegraphics[width=\textwidth]{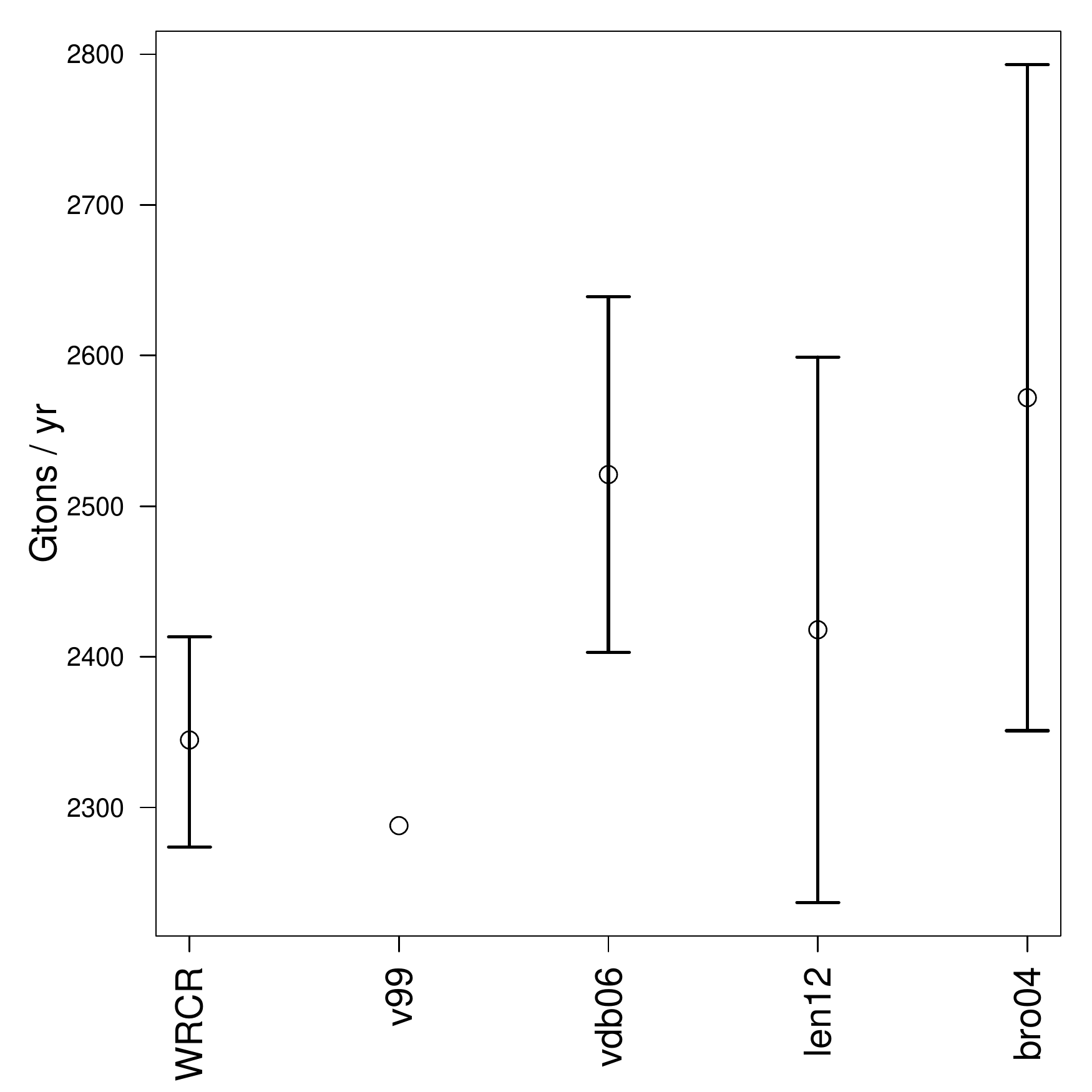}
      \subcaption{SMB over all ice sheets}
   \end{subfigure}
  \begin{subfigure}[b]{.4\textwidth}
     \includegraphics[width=\textwidth]{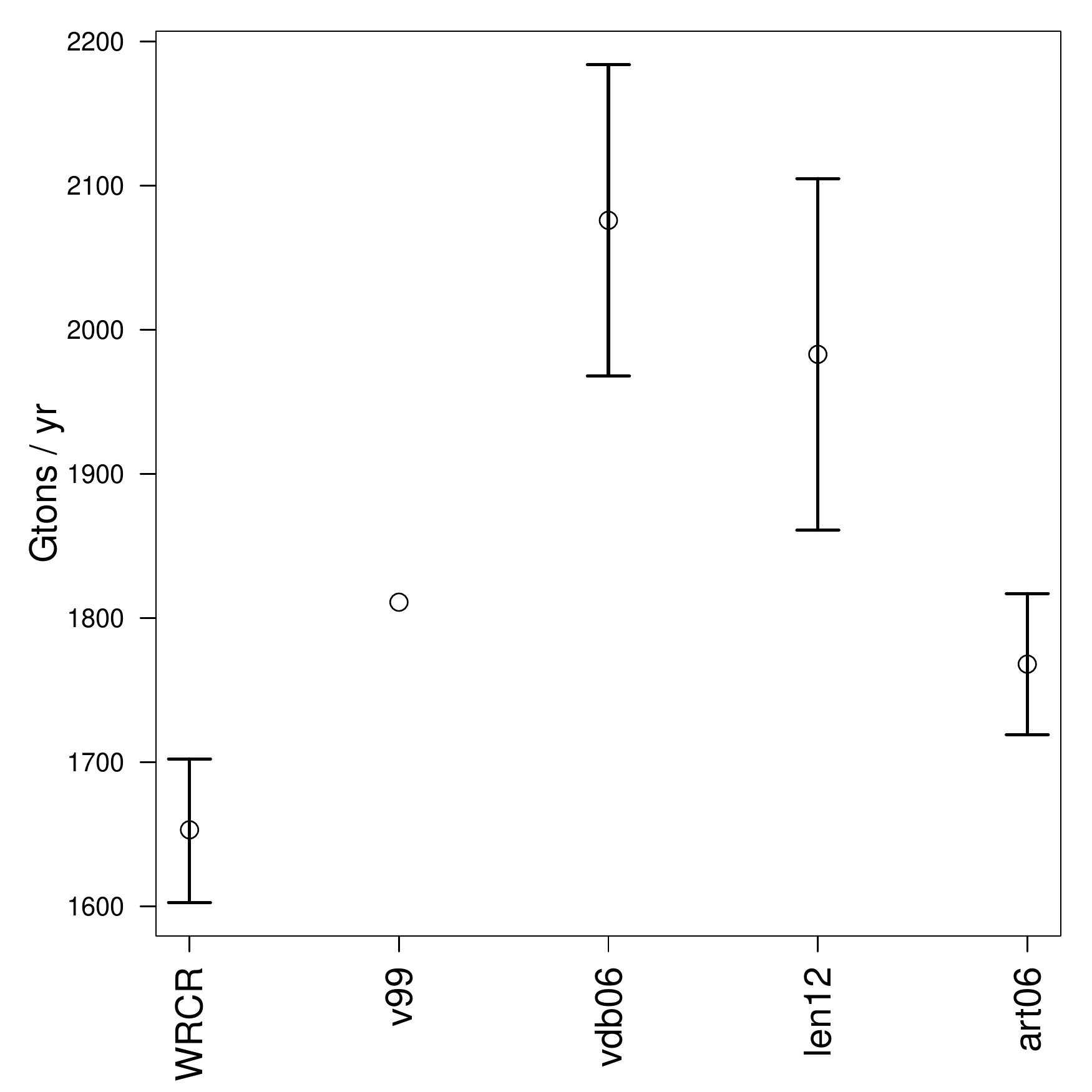}
           \subcaption{SMB over grounded ice sheets}
   \end{subfigure}
  \end{center} 
  \vspace{-2mm}
       \caption{(a) Comparison between our estimate of net SMB over all Antarctic ice sheets and previous estimates. Notably, \citet{vaughan1999} only provides a point estimate. Note that our interval estimate (WRCR) overlaps other published interval estimates (e.g. \citet{vaughan1999} (v99), \citet{lenaerts2012} (len12), \citet{vandeberg2006} (vdb06), and \citet{bromwich2004} (bro04)). (b) Comparison between our estimate of SMB over Antarctica's grounded ice sheets and previous estimates. Note that our estimate is lower than most previous estimates on average.} \label{fig:comp}
\end{figure}

Our point estimate for net SMB over all Antarctic ice sheets, 173 mm w.e. $\textrm{yr}^{-1}$, is lower than most previously estimated values (see, e.g., \citet{bromwich2004}, \citet{vandeberg2006}, and \citet{lenaerts2012}) but exceeds \citet{vaughan1999}. Our prediction intervals, however, intersect those of other estimates (see Figure \ref{fig:comp}). Over the grounded ice sheet, our estimate of total SMB is significantly lower than previous analyses \citep{vaughan1999, vandeberg2006, arthern2006, bromwich2004, lenaerts2012}.
\vspace{-2mm}

It is important to note that several previous estimates have point-wise SMB predictions higher than has been observed by field measurements. For example, \citet{lenaerts2012} and \citet{vandeberg2006} predicted SMB values as high as 4000 mm w.e. $\textrm{yr}^{-1}$ at some locations, while the highest SMB field measurement is 2860 mm w.e. $\textrm{yr}^{-1}$, and 1665 mm w.e. $\textrm{yr}^{-1}$ is the highest A-rated SMB field measurements. Additionally, many of the highest SMB measurements were taken using less reliable methods, and our model more heavily weights more reliable data. While SMB values higher than those recorded in measurements may certainly exist, these predicted values are about 40\% greater than any recorded value, even when compared to in situ measurements known to be in high accumulation regions. Furthermore, previous estimates rarely have point-wise estimates below 0 mm w.e. $\textrm{yr}^{-1}$ \citep{vaughan1999,arthern2006,bromwich2004,lenaerts2012}; however, there are field measurements as low as -317 mm w.e. $\textrm{yr}^{-1}$ (refer to the histogram in Figure \ref{fig:alldat}). This could partially explain why our estimate for net surface mass balance is lower than other estimates. Since our model is based on the most recent and complete data compilation \citep{favier2013} with data quality ratings \citep{magand2007} explicitly accounted for, our analyses are given the greatest opportunity for accurate estimation of both SMB and the associated uncertainty. Lastly, because our estimate comes from a statistical model, rather than partial differential equation climate models \citep{bromwich2004,lenaerts2012}, it is more firmly bound to field measurements and is not deterministic. However, in future analyses we propose using methods that synthesize both data types.

Because our model is designed to incorporate data from many sources and account for various types of uncertainty, this model is adaptable as new data become available. Thus, as newer data are available, our estimates for net SMB, spatial prediction uncertainty, and field measurement proposals will be updated accordingly. A future goal of this model is to extend the spatial method described to a spatiotemporal model that can be used to assess trends in Antarctic SMB spatially and temporally. While our dataset consists of time-averaged SMB measurements, there is implicitly a time-series at every location in the dataset. Temporal analyses have been done on a small number of ice cores. For example, \cite{thomas2017} utilized 80 ice core sites to analyze temporal trends in SMB and argued that, generally, SMB has not decreased over the past 1000 years.  Using available time-series data, we will be able to explore how net SMB and prediction uncertainty has changed in time, as well as identify temporal trends over space. Additionally, we plan to utilize climate model output and remote sensing data with field measurements through the framework of a computer model \citep{higdon2008,ranjan2011}, as discussed in Section \ref{sec:extension}.

\section*{Acknowledgments}

The authors gratefully acknowledge NASA for support provided by award \#NNX16AQ61G to Summer Rupper, Shane Reese, and William Christensen.

\bibliographystyle{apacite}
\bibliography{refs}

\newpage
\vspace{-2mm}
\begin{table}[H]
\centering
\scriptsize
\begin{tabular}{rrrrrrrr}
  \hline
 Model & VST  & NNGP Specification & Latent Variable & Covariance & PRMSE & 90\% Coverage & CRPS \\ 
  \hline
  1 & No & None & None & None & 122.95 & 0.95 & 64.71 \\ 
  2 & Yes & None & None & None & 127.01 & \textbf{0.93} & 62.55 \\ 
  3 & Yes & Univariate & None & Spherical & \textbf{69.57} & 0.95 & 32.02 \\ 
  4 & Yes & Univariate & None & Euclidean & 117.59 & 0.94 & 59.96 \\ 
  5 & Yes & Univariate & None & Separable & 71.20 & 0.94 & 31.84 \\ 
  6 & Yes & Univariate & None & Non-separable & 71.51 & 0.96 & 32.94 \\ 
  7 & Yes & Univariate & Yes & Spherical & \textbf{70.48} & 0.96 & 33.87 \\ 
  8 & Yes & Univariate & Yes & Euclidean & 116.51 & 0.94 & 58.78 \\ 
  9 & Yes & Univariate & Yes & Separable & 71.21 & 0.96 & 33.01 \\ 
  10 & Yes & Univariate & Yes & Non-separable & 72.56 & 0.97 & 34.82 \\ 
  11 & Yes & Multivariate & None & Spherical & 131.27 & 0.86 & 56.80 \\ 
  12 & Yes & Multivariate & None & Euclidean & 117.60 & 0.94 & 54.34 \\ 
  13 & Yes & Multivariate & None & Separable & 73.27 & \textbf{0.93 }& \textbf{31.20} \\ 
  14 & Yes & Multivariate & None & Non-separable & 71.45 & 0.94 & \textbf{30.16} \\ 
  15 & Yes & Multivariate & Yes & Spherical & 126.78 & 0.80 & 55.99 \\ 
   16 & Yes & Multivariate & Yes & Euclidean & 116.82 & 0.94 & 54.15 \\ 
   17 & Yes & Multivariate & Yes & Separable & 73.24 & 0.94 & \textbf{31.20} \\ 
 \textbf{18} & Yes & Multivariate & Yes & Non-separable & 71.45 & \textbf{0.93} & \textbf{30.08} \\ 
 19 & No & Univariate & None & Spherical & 135.20 & \textbf{0.90} & 67.61 \\ 
 20 & No & Univariate & None & Euclidean & 114.04 & 0.97 & 63.78 \\ 
 21 & No & Univariate & None & Separable & 71.41 & 0.96 & 34.90 \\ 
 22 & No & Univariate & None & Non-separable & \textbf{70.79} & 0.96 & 34.80 \\ 
 23 & No & Univariate & Yes & Spherical & 139.73 & 0.83 & 69.29 \\ 
 24 & No & Univariate & Yes & Euclidean & 112.37 & 0.97 & 61.90 \\ 
 25 & No & Univariate & Yes & Separable & 71.20 & 0.97 & 38.39 \\ 
 26 & No & Univariate & Yes & Non-separable & \textbf{70.63} & 0.97 & 37.57 \\ 
 27 & No & Multivariate & None & Spherical & 128.50 & \textbf{0.89} & 59.49 \\ 
 28 & No & Multivariate & None & Euclidean & 116.35 & 0.97 & 58.11 \\ 
 29 & No & Multivariate & None & Separable & 74.17 & 0.95 & 33.21 \\ 
 30 & No & Multivariate & None & Non-separable & 71.66 & 0.95 & 31.72 \\ 
 31 & No & Multivariate & Yes & Spherical & 125.98 & 0.84 & 58.92 \\ 
 32 & No & Multivariate & Yes & Euclidean & 115.32 & 0.97 & 57.99 \\ 
 33 & No & Multivariate & Yes & Separable & 72.99 & 0.96 & 34.34 \\ 
 34 & No & Multivariate & Yes & Non-separable & 71.65 & 0.95 & 31.76 \\ 
   \hline
\end{tabular}
\caption{Model comparison for various models considered. Best performances are bolded (i.e. coverage closest to 90\%, lowest PRMSE, and lowest CRPS). }\label{tab:modelcomp}
\end{table}

\newpage

 \vspace{-4mm}
  \begin{table}[H] 
\footnotesize
\begin{center}
\begin{tabular}{rrrrrr}
& Estimate & 95\% Credible Interval & Units  \\ 
  \hline
  \bf{All Antarctic Ice Sheets} & & & \\
Net SMB &  2345 & (2273 , 2413)&  $\textrm{Gton} \, \textrm{yr}^{-1}$ \\ 
Average SMB  & 173 & (168 , 178) & mm w.e. $\textrm{yr}^{-1}$ \\ 
  \bf{Grounded Ice Sheets} &  & & \\
Net SMB  &  1653 & (1603 , 1702) &  $\textrm{Gton} \, \textrm{yr}^{-1}$ \\ 
Average SMB  & 139 & (134 , 143) & mm w.e. $\textrm{yr}^{-1}$ \\ 
\end{tabular}
\end{center}
\vspace{-2mm}
\caption[Net and average SMB estimates]{Net and average SMB over all ice sheets and grounded ice sheets. Both point estimates and 95\% credible interval results are given.} \label{tab:SMB}
\end{table} 

\newpage

\begin{table}[H] 
\scriptsize
\centering
\begin{tabular}{rrrrrrrrrr}
 & Covariate &Mean & Mode & std. dev. & 95\% credible interval & Geweke $z$-score  & HW $p$-value \\ 
  \hline
  $V_{11}$& & 0.177 & 0.180 & 0.013 & (0.157 , 0.206) & -1.079  & 0.280 \\
    $V_{22}$& &  0.203 & 0.209 & 0.026 & (0.164 , 0.262) & -1.010  &  0.093 \\
  $V_{33}$& & 0.077 & 0.075 & 0.009 & (0.055 , 0.092) & -0.185 & 0.266 \\
  $V_{44}$& & 0.438 & 0.457 & 0.045 & (0.377 , 0.551) & 1.307 & 0.068  \\
  $V_{12}$& & 0.058 & 0.059 & 0.012 & (0.036 , 0.082) & 0.916 & 0.426  \\
  $V_{13}$& & -0.016 & -0.017 & 0.007 & (-0.030 , -0.004) & 0.320 &   0.479 \\
  $V_{14}$& &  0.176 & 0.177 & 0.017 & (0.146 , 0.211) & -0.366 &  0.314 \\ 
  $V_{23}$& &  0.109 & 0.111 & 0.012 & (0.089 , 0.134) & -0.941 & 0.674 \\
  $V_{24}$& & 0.277 & 0.289 & 0.033 & (0.228 , 0.354) & 0.634 & 0.613  \\ 
  $V_{34}$& & 0.123 & 0.126 & 0.016 & (0.095 , 0.157) & -1.062 & 0.073 \\
$\rho_1$ & & 0.098 & 0.112 & 0.060 & (0.025 , 0.238) & -1.334 & 0.225   \\
$\rho_2$ & & 0.406 & 0.421 & 0.059 & (0.295 , 0.521) & -0.164 & 0.228 \\
$\alpha$ & & 0.271 & 0.248 & 0.015 & (0.248 , 0.297) & -1.129 & 0.612 \\
$\delta$ & & 0.393 & 0.404 & 0.166 & (0.068, 0.653) & -0.798 & 0.106 \\
$\nu$ & & 0.455 & 0.472 & 0.350 & (0.041 , 0.970) & -0.413 & 0.174 \\
$\tau_A^2$ & & 0.058 & 0.058 & 0.002 & (0.054 , 0.063) & -1.659 & 0.113   \\
$\tau_B^2$ & & 0.062 & 0.062 & 0.003 & (0.057 , 0.068) & -0.627 &  0.405 \\
$\tau_C^2$ & & 0.065 & 0.066 & 0.004 & (0.060 , 0.073) & -0.375 &   0.580 \\
$\theta$  & & 0.500 & 0.500 & 0.016 & (0.471 , 0.532) & -1.172 & 0.334 \\
$\beta_1$ & el & 0.125 & 0.103 & 0.070 & (-0.037  , 0.239) & -1.174 & 0.083  \\
$\beta_2$ & dc & -0.109 & -0.114 & 0.048 & (-0.202 , -0.016) & -1.648  & 0.178   \\
$\beta_3$ & temp & 0.479 & 0.463 & 0.086 & (0.306 , 0.635) & -1.347 & 0.166  \\
$\beta_4$ & el$\times$dc& -0.096 & -0.088 & 0.052 & (-0.186 , 0.019) & -0.996 & 0.173   \\
$\beta_5$ &  el$\times$temp & 0.014 & 0.012 & 0.021 & (-0.029 , 0.053) & -1.770 &  0.430 \\
$\beta_6$ & dc$\times$temp & -0.304 & -0.312 & 0.043 & (-0.396 , -0.226) & 0.568 &  0.262 \\
$\beta_7$ & el$\times$dc$\times$temp & 0.043 & 0.043 & 0.019 & (0.008 , 0.081) & -0.626 & 0.292   \\
\end{tabular}
\caption{Posterior distribution summary statistics, including posterior mean, standard deviation, and 95\% credible intervals. Geweke ($z$-scores) and Heidelberger-Welch (HW p-value) convergence diagnostics for all model parameters are given.}\label{tab:parameter}
\end{table}
\vspace{-3mm}

\newpage

\appendix

\vspace{-3mm}
\section{Gibbs Sampling for Final Model}
\label{sec:Gibbs}

The model is described in Section \ref{sec:mod}. Let $\bD_\tau$ be a diagonal matrix with the current values of $\tau^2_{y^*_{ij}}$ (the error associated with current value of $y^*_{ij}$). Given our prior distributions on hyperparameters $\tau^2_A$, $\tau^2_B$, $\tau^2_C$, $\theta$, $\bbeta$, $\bV$, $\rho_1$, $\rho_2$, $\nu$, $\alpha$, and $\delta$:
\begin{equation}
\begin{aligned}[c]
  \tau_A^2 &\sim \text{IG}(a_{A},b_{A}), \\
  \tau_B^2 &\sim \text{IG}(a_{B},b_{B}), \\
  \tau_C^2 &\sim \text{IG}(a_{C},b_{C}), \\
  \theta &\sim \text{Beta}(a_\theta,b_\theta),
    \end{aligned}
    \qquad
    \begin{aligned}[c]   
  \bbeta &\sim \mathcal{N}( \bm_\beta, \bV_\beta ), \\
      \rho_1 &\sim \text{Gamma}(a_{\rho_1},b_{\rho_1}), \\
       \rho_2 &\sim \text{Gamma}(a_{\rho_2},b_{\rho_2}), \\
\bV &\sim \text{IW}(\bI,p+1),
  \end{aligned}
      \qquad
  \begin{aligned}[c]
  \alpha &\sim \text{Unif}[0,2], \\
  \nu &\sim \text{Unif}[0,1], \\
  \delta &\sim \text{Gamma}(a_\delta,b_\delta), \\
    \end{aligned}
  \end{equation}
where $a_A=20$, $b_A=6$, $a_B=20$, $b_B=8$, $a_C=20$, $b_C=10$, $a_\theta=1$, $b_\theta=1$, $\bm_\beta = \bzero$, $\bV_\beta = \bI$, $a_{\rho_1} = 2$, $b_{\rho_1} = 20$, $a_{\rho_2} = 1$, $b_{\rho_2} = 10$, $a_\delta = 1$, $b_\delta = 1$, and the Gamma and Inverse Gamma distributions use the shape-rate parameterization.
The full conditional distributions, which we denote $\cdot | \cdots$, are
\begin{align*}
\bbeta | \cdots &\sim \mathcal{N}( V_\beta^* m_\beta^*,V_\beta^*) \\
\tau^2_A | \cdots &\sim IG(a_{\tau_A}^*,b_{\tau_A}^*) \\
\tau^2_B | \cdots &\sim IG(a_{\tau_B}^*,b_{\tau_B}^*) \\
\tau^2_C| \cdots &\sim IG(a_{\tau_C}^*,b_{\tau_C}^*) \\
V| \cdots &\sim \text{IW}(a_V^*,b_V^*) \\
\bw_i | \cdots &\sim \mathcal{N}( V_{w_i}^* m_{w_i}^*,V_{w_i}^*)\\
y^*_{ij} \neq A | \cdots &\sim Bern(p_{y_{ij}}^*) \\
\theta | \cdots &\sim Beta(a_\theta^*,b_\theta^*)
\end{align*} 
where
{ \scriptsize
\begin{align*}
    V_\beta^* &= \left( \bX^T \bD_\tau^{-1} \bX  + V_\beta^{-1} \right)^{-1} \\
    m_\beta^* &= V_\beta^{-1} m_\beta + \sum_i \sum_j \bx_{ij} ( y_{ij} - w_i) / \tau_{y^*_{ij}}^2  \\
    V_{w_i}^* &= \left( \sum_{j:j \in s_i} \bz_{ij} \bz_{ij}^T/\tau_{y^*_{ij}}^2 + F_{s_i}^{-1} + \sum_{t:U(s_i)}  B^T_{t,s_i} F_t^{-1} B_{t,s_i}   \right)^{-1}\\
  m_{w_i}^* &= \sum_{j: j \text{ at } s_i} \bz_{ij} (y_{ij} - \bx_{ij}^T \bbeta ) /\tau_{y^*_{ij}}^2 + F_{s_i}^{-1} B_{s_i} \bw_{N(s_i)} + \sum_{t:U(s_i)}  B^T_{t,s_i} F_t^{-1} \ba_{t,s_i} \\
  a_\theta^*  &= a_\theta + \sum_i \sum_j \bone(y^*_{ij} =B) \\
b_\theta^* &= b_\theta + \sum_i \sum_j \bone(y^*_{ij} =C) \\  
a_V^* &= p + 1 + N_u\\
b_V^* &= \Psi^{-1} +  \sum_i  (\bw_{i} - B_{s_i} {\bw_{N(s_i)}} ) (\bw_{i} - B_{s_i} {\bw_{N(s_i)}} )^T /R_{s_i} \\
a_{\tau_A}^* &= a_\tau + \frac{1}{2}\sum_i \sum_j \bone(y^*_{ij} = A ) \\
b_{\tau_A}^* &= b_\tau + \frac{1}{2} \sum_i \sum_j (y_{ij} - \bx_{ij}^T \bbeta - w_i)^2 \bone(y^*_{ij} = A )  \\
p_{y_{ij}^*} &=  \frac{\theta N(y_{ij} | w_i + \bx_i^T \bbeta, \tau_B^2) }{\theta N(y_{ij} | w_i + \bx_i^T \bbeta, \tau_B^2) + (1-\theta) N(y_{ij} | w_i + \bx_i^T \bbeta, \tau_C^2)} \\
\end{align*}
}
and $\ba_{t,s_i}$ is as it is defined in \citet{datta2016a}. Parameters $a_{\tau_B}^*$, $b_{\tau_B}^*$, $a_{\tau_C}^*$, and $b_{\tau_C}^*$ are updated similarly to $a_{\tau_A}^*$ and $b_{\tau_A}^*$.

%
\end{document}